\documentclass{aa}
\usepackage{graphicx}
\usepackage{natbib}
\usepackage{txfonts}

\begin{document}

 \title{Modifications of thick-target model: re-acceleration of electron beams by static and stochastic electric fields}

   \titlerunning{Modifications of thick-target model: re-acceleration
     of electron beams ... }
   \authorrunning{M.~Varady et al.}

   \author{M. Varady
          \inst{1}
          \and
          M. Karlick\'{y}
          \inst{2}
          \and
          Z. Moravec \inst{1}
          \and
          J. Ka\v{s}parov\'{a}\inst{2}
         }


    \offprints{Varady}

   \institute{ J.E. Purkyn\v{e} University, Physics Department, \v{C}esk\'{e}
              ml\'{a}de\v{z}e 8, 400 96 \'{U}st\'{\i} nad Labem, Czech
              Republic, \\
              \email{mvarady@physics.ujep.cz}
         \and
         Astronomical Institute of the Academy of Sciences of the Czech
              Republic, v.v.i., 25165 Ond\v{r}ejov, Czech Republic \\
             \email{karlicky@asu.cas.cz}}

   \date{Received 29 July 2013 / Accepted 21 November 2013}


\abstract
{The collisional thick-target model (CTTM) of the impulsive phase of solar
flares, together with the famous CSHKP model, presented for many years a
``standard'' model, which straightforwardly explained many observational
aspects of flares. On the other hand, many critical issues
appear when the concept
is scrutinised theoretically or with the new
generation of  hard X-ray (HXR) observations. The famous ``electron
number problem'' or problems related to transport of enormous
particle fluxes though the corona represent only two of them. To resolve
the discrepancies, several modifications of the CTTM appeared.}
{ We study two of them based on the global and local re-acceleration of
non-thermal electrons by static and stochastic electric fields
during their transport from the coronal acceleration site to the
thick-target region in the chromosphere.
We concentrate on a comparison of the non-thermal electron distribution
  functions, chromospheric energy deposits, and HXR spectra
  obtained for both considered modifications with the CTTM itself.}
{The results were obtained using a relativistic test-particle
approach. We simulated the transport of non-thermal electrons
with a power-law spectrum including the
influence of scattering, energy losses, magnetic mirroring, and
also the effects of the electric fields corresponding to
both modifications of the CTTM.}
{We show that both modifications of the CTTM
    change the outcome of the chromospheric bombardment
in several aspects. The modifications lead
    to an increase in chromospheric energy deposit, change of its
    spatial distribution, and a substantial increase in the corresponding HXR
    spectrum intensity.}
{The re-acceleration in both models
  reduces the demands on the efficiency of the primary coronal accelerator,
  on the electron fluxes transported from the corona downwards, and on the total
number of accelerated coronal electrons during flares.}


   \maketitle

\section{Introduction}

The CTTM of the  impulsive phase of solar
flares \citep{brown1971} for many years presented a successful tool
not only for interpreting the processes related to the energy deposition
and HXR production in the footpoint regions of flare
loops, but also for naturally explaining many other observational aspects of flares
like the Neupert effect \citep{denis1993},  the time correlation of
footpoint HXR intensity and intensities of
chromospheric lines \citep{radziszewski2007, radziszewski2011}, or
the radio signatures of particle transport from the corona towards the
chromosphere \citep{bastian1998}.
Nevertheless, especially with the onset of modern HXR
observations such as Yohkoh/HXT, RHESSI
\citep{kosugi1991, lin2002}, a continuously growing number of discrepancies with
the CTTM were beginning to appear. The most striking one is the old standing
problem concerning the very high electron fluxes required to explain the
observed high HXR footpoint intensities. This problem is particularly
acute in the context of the ``standard'' CSHKP flare model when assuming a
single coronal acceleration site \citep{sturrock1968, kopp1976, shibata1996},
where enormous numbers of electrons involved in the impulsive phase have to be
gathered, accelerated, and then transported to the thick-target region located
 in the chromosphere \citep{brown_melrose1977, brown2009}. Another serious
class of problems appears as a consequence of enormous electric currents
arising from the transport of high electron fluxes through the corona down to
the chromosphere and the inevitable  generation of the neutralising return
current \citep{oord1990,matthews1996,karlicky2009,holman2012}. Also the
recent measurements of the vertical extent of chromospheric HXR sources
\citep{battaglia2012} are inconsistent with the values predicted by the CTTM.

Generally, it is very difficult to explain energy transport by means of
electron beams with enormous fluxes from the primary coronal
acceleration sites assumed to be located in highly structured coronal current
sheets \citep{shibata2001,barta2011b,barta2011a} to the thermalisation regions
that lie relatively deep in the atmosphere and that produce the observed intensities
of footpoint HXR emission in the frame of classical CTTM. Therefore various
modifications of the CTTM have been proposed to solve the problems.
\citet{fletcher2008} suggest a new mechanism of energy transport from the
corona downwards by  Alfv\'{e}n waves, which in the chromosphere accelerate
electrons to energies for X-ray emission. Furthermore, \citet{KarlickyKontar12}
have investigated an electron acceleration in the beam-plasma system.
Despite efficient beam energy losses to the
thermal plasma, they have found that
a noticeable part of the electron population is accelerated by
Langmuir waves produced in this system. Thus, the electrons accelerated during
the beam propagation downwards to the chromosphere can reduce the beam flux in
the beam acceleration site in the corona requested for X-ray emission.
Another modification of the CTTM is the local re-acceleration thick-target
model (LRTTM) that has been suggested by \citet{brown2009}. The model assumes a
primary acceleration of electrons in the corona and their transport along the
magnetic field lines downwards to the thick-target region. Here they are
subject to secondary local re-acceleration by stochastic electric fields
generated in the stochastic current sheet cascades \citep{turkmani2005,
turkmani2006} excited by random photospheric motions.

\citet{karlicky1995} studied another idea -- the global
re-acceleration thick-target model (GRTTM). The
beam electrons accelerated in the primary coronal acceleration site are
on their path from  the corona to the chromosphere constantly
re-accelerated. Such a re-acceleration is caused by small
static electric fields generated by the
electric currents originating due to the helicity
of the magnetic field lines forming the flare loop
\citep[{e.g.}][]{gordovskyy2011, gord2012, gord2013}. The magnitude of the
static electric field reaches its maximum in the thick-target region
owing to the sharp decrease in electric conductivity in the
chromosphere and to the prospective convergence of magnetic field in this
region.

In this paper we study the effects of the local and global re-acceleration of beam electrons
at locations close to the hard X-ray chromospheric sources.
Section~\ref{sec:des.model} describes our approximations of LRTTM  and GRTTM
and their implementation to a relativistic test-particle code.
In Section~\ref{sec:results} we compare both modifications with CTTM in terms of
electron beam distribution functions,  chromospheric energy deposits, and HXR spectra.
Modelled HXR spectra are also forward-fitted to obtain beam parameters
under the assumption of pure CTTM regardless of any re-acceleration.
The results are summarised and discussed in Section~\ref{sec:concl}.

\section{Model description}\label{sec:des.model}

\subsection{Beam properties and target atmosphere}

The simulations presented in this work start with an injection of
an initial electron beam into a closed magnetic loop at its
summit point using a test-particle approach \citep{varady2010}.
Physically, the initial beam represents a population of non-thermal
electrons generated at the primary acceleration site located in the
corona above the flare loop. Our simulations do not treat the primary
acceleration itself. The non-thermal electrons are assumed to obey a
single power law in energy, so their initial spectrum (in units:
\mbox{electrons~cm$^{-2}$~s$^{-1}$~keV$^{-1}$}) is
\begin{eqnarray}
F(E, z_0) = \left\{\begin{array}{ll}
(\delta_{\mathrm{p}}-2)\frac{\mathcal{F}_0}{E_0^2}\left(\frac{E}{E_0}\right)^{-\delta_{\mathrm{p}}}
& , \ \ \  \mbox{for} \ \ \  E_0 \le E \le E_1   \\
0  & ,  \  \ \  \mbox{for other} \ E   \\
\end{array} \right.  \ \
\end{eqnarray}
\citep{nagai1984}.
The electron flux at the loop top, which corresponds to the column density
$z_0=0$, is determined by the total energy flux
$\mathcal{F}_{\mathrm{0}}$, the low and high-energy
cutoffs $E_0$, $E_1$ and the power-law index $\delta_{\mathrm{p}}$. All
the models presented in this work start with the same initial beam
parameters $\delta_{\mathrm{p}} =3$, $E_0=10$~keV and $E_1=400$~keV.
For $\mathcal{F}_{\mathrm{0}}$ we use two values
$\mathcal{F}_0 = 5\times10^9$ and $10^{11}$~erg\,cm$^{-2}$\,s$^{-1}$, with
the latter only as the CTTM reference flux for a comparison with the
  models of secondary re-acceleration.

We study two various cases of initial pitch angle distribution. The
pitch angle $\vartheta$ determines the angle between the non-thermal
electron velocity component parallel to the magnetic field line
$\varv_{\parallel}$ and the total electron velocity $\varv$
\begin{equation}
\mu \equiv \cos{\vartheta} = \frac{\varv_{\parallel}}{\varv} \ .
\end{equation}
The initial $\mu$-distribution is given by function $M(\mu_0)$ and
must be normalised. The angularly dependent initial electron flux is then
\begin{equation}
F(E, \mu_0, z_0) = M(\mu_0) F(E, z_0) \ , \qquad \int_{-1}^{1}M(\mu_0)\mathrm{d}{\mu_0} = 1\ .
\end{equation}
We consider two extreme cases:
\begin{enumerate}
\item a fully focussed beam
   \begin{equation}
       M^\mathrm{FF}\equiv
       M(\mu_0)=\frac{1}{2}\delta(\mu_0-\mu_{\mathrm{c}}) \ ,
     \label{eq:case_A}
   \end{equation}
   where $\delta$ is the Dirac function and $\mu_{\mathrm{c}}=\pm1$; and
\item a semi-uniformly distributed beam
\begin{equation}
M^\mathrm{SU}\equiv M(\mu_0) = \left\{ \begin{array}{r@{\quad}l}
    1, & \mu_0  \in(-1, -0.5) \cup (0.5, 1) \\
    0, & \mu_0\in(-0.5, 0.5)  \ .  \\ \end{array} \right.
\label{eq:case_B}
\end{equation}
\end{enumerate}
The initial pitch angle distribution
reflects the properties of the primary
coronal accelerator.
The first distribution may represent an extreme case of an electron beam
accelerated in the coronal current sheet with an X-point, and the second
is close to the outcome of the acceleration mechanisms involving the
plasma wave turbulence in a second-order Fermi process \citep{winter2011}.
The electrons with negative $\mu_0$ propagate to the left,
with positive $\mu_0$ to the right half of the loop. Since we study
the effects of the electron beam bombardment
of the chromosphere, we excluded the population with $\mu_0\in(-0.5,
0.5)$ from the uniform distribution. This approximation
substantially decreases the computational cost. The choice of
$M(\mu_0)$ influences the initial energy flux along magnetic
field lines towards a single left or right footpoint. The parallel
fluxes towards individual footpoints are $\mathcal{F}_0/2$ for
$M^\mathrm{FF}$ and $3\mathcal{F}_0/8$ for $M^\mathrm{SU}$,
respectively. The total number of non-thermal electrons injected
into the loop per  unit area and  time is
$\approx$1.6$\times10^{17}$~electrons\,cm$^{-2}$\,s$^{-1}$
  (relevant to the energy flux $\mathcal{F}_0 = 5\times10^9$~erg\,cm$^{-2}$\,s$^{-1}$
and both pitch angle distributions).

\begin{figure*}[!t]
  \centerline{\includegraphics[width=8.2cm]{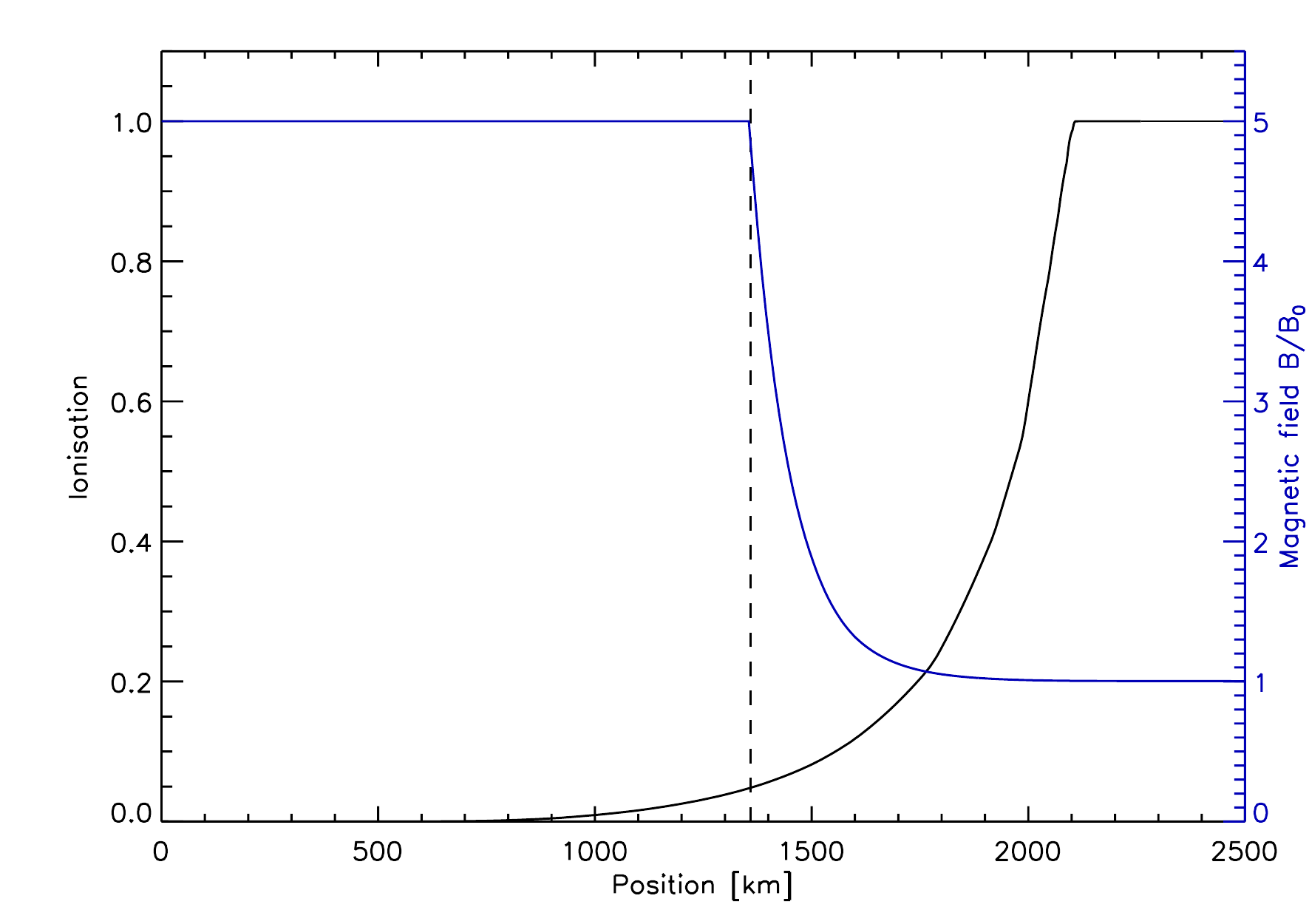}
              \includegraphics[width=8.2cm]{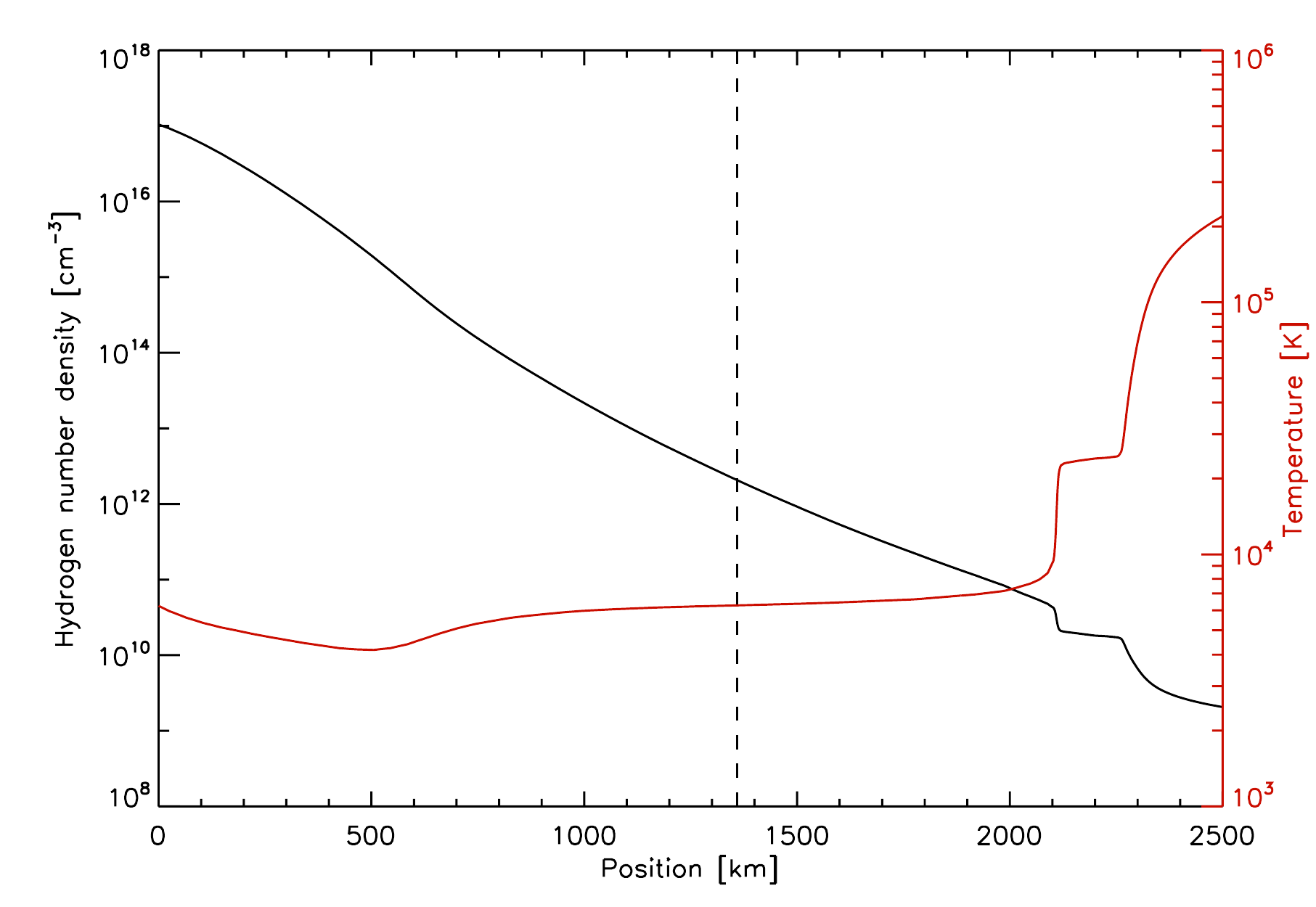}}
\caption{{\it Left:} Hydrogen ionisation (black line) and relative magnetic field
  strength $B/B_{\mathrm{0}}$ (blue line), {\it right} temperature
  (red line), and hydrogen density (black line) in the lower parts of the VAL C
  atmosphere. The dashed line indicates the lower boundary of the magnetic mirror.}
\label{fig:1}
\end{figure*}

We consider a converging magnetic field along the loop towards the
photosphere with a constant mirror ratio $R_{\mathrm{m}} \equiv
B_1/B_0=5$, where $B_0$
and $B_1$ are the magnetic fields at the loop top in the corona and at
the base of the loop in the photosphere, respectively. To model the field
convergence we adopted the formula proposed by \citet{bai1982},
where the magnetic field strength $B$ is only a function of the column
density $z$ calculated from the loop top downwards
\begin{eqnarray}
\frac{B(z)}{B_{\mathrm{0}}} = \left\{\begin{array}{ll}
1 + (R_{\mathrm{m}} - 1)(z/z_{\mathrm{m}})^2   & , \ \ \  \mbox{for} \ \ \  z \le z_{\mathrm{m}}  \\
R_{\mathrm{m}}  & ,  \  \ \  \mbox{for} \ \ \  z \ge z_{\mathrm{m}}  \\
\end{array} \right.  \ \ ,
\end{eqnarray}
where $z_{\mathrm{m}}=4\times10^{19}$~cm$^{-2}$. For the VAL~C
atmosphere \citep{VAL1981} $z_{\mathrm{m}}$ is located in the
chromosphere -- corresponding position $s_{\mathrm{m}}=1.36$~Mm,
temperature $T_{\mathrm{m}}=6270$~K and density $n_{\mathrm{m}}=2\times10^{12}$~cm$^{-3}$.
The adopted configuration of the magnetic field is shown in
Fig.~\ref{fig:1}. The convergence of the magnetic field in
the vicinity of the loop footpoints influences the model in two
aspects. First, only part of the beam electrons with low pitch angles satisfying
the condition $\sin^2\vartheta \le \frac{1}{R_{\mathrm{m}}}$ passes
through the magnetic mirror. Second, the corresponding flux is
focussed thanks to the field convergence that results in an increase
in the energy deposit per unit volume in the constricted
flux tube. The remaining beam particles are reflected by the
mirror and move back to the loop top and further to the second part of
the loop \citep{Karlicky&Henoux1993}.

The corresponding energy deposits, non-thermal electron distribution
functions, and the HXR spectra are determined primarily by the parameters of
the electron beam itself, but also by the properties of the target atmosphere.
The results are obtained for the VAL~C atmosphere (see
Fig.~\ref{fig:1}) \citep{VAL1981}, which was extrapolated to the hot
$\sim$1~MK and low density $10^8$~--~$10^9$~cm$^{-3}$ corona. The
length of the whole loop is $L=20$~Mm, so the source of the energetic
particles (primary coronal acceleration site) is located at $s=10$~Mm.

The hydrodynamic flare models show that a rapid and
massive flare energy release in the thick-target
region dramatically changes the temperature and ionisation
structure in the chromosphere
on very short timescales $\le 1$~s \citep{abbett1999, allred2005, kasparova2009}.
Therefore it also influences the thermalisation rate of the
non-thermal electrons \citep{emslie1978, kasparova2009} and thus
the outcome of the bombardment \citep{varady2013}.
Using a hydrodynamic flare code combined with a test-particle code
\citep{varady2010}, we tested the influence in increased temperature and
change of ionisation due to the flare heating  on the
HXR spectra produced in the thick-target region and on the corresponding energy
deposits. We found only relatively minor changes in comparison with the
results for the quiet VAL~C atmosphere. Therefore
only results for the quiet VAL~C  atmosphere are presented in
this study.

\subsection{Test-particle approach}

The problem of collisional particle transport in a partially ionised atmosphere
in the cold target approximation was analysed by \citet{emslie1978}.
\citet{bai1982} presented a Monte-Carlo method that is useful for  computer
implementation of the transport of energetic electrons in a fully ionised hydrogen
plasma in a non-uniform magnetic field. It has been shown by
\citet{mackinnon1991} that the coupled system of
stochastic equations presented in \citet{bai1982} is formally equivalent to the
corresponding Fokker-Planck (FP) equation, therefore the method proposed by
\citet{bai1982} has to give equivalent results as the direct solution of the
FP equation. We
modified the approach of \citet{bai1982}  for a partially
ionised cold target and developed a relativistic test-particle code. The
code follows the motion of a chain of beam electron clusters, test-particles
with a power-law spectrum along a magnetic field line described
with the following equation of motion
\begin{equation}
\frac{\mathrm{d}\vec{p}_\mathrm{e}}{\mathrm{d}t} = -
{\vec{C}}_{\mathrm{e}}(\varv_{\mathrm{e}}) + {\vec{F}}_{\mathrm{m}} -
e{\vec{E}} \ ,
\label{eq:pohybovka}
\end{equation}
where $\vec{p}_{\mathrm{e}}$ is the momentum of the electron cluster,
$-\vec{C}_{\mathrm{e}}(\varv_{\mathrm{e}})$ is the collisional drag also responsible for the
effects of scattering, $\vec{F}_{\mathrm{m}}$ is the magnetic mirror force, and the term
$-e\vec{E}$ expresses the force controlling the secondary acceleration.

\subsection{Collisional thick-target model -- CTTM}

In the scenario of classical CTTM, the non-thermal electrons lose their energy
and are scattered by the Coulomb collisions with the particles of the
ambient plasma (see the  term $-\vec{C}_{\mathrm{e}}(\varv_{\mathrm{e}})$
  in equation~(\ref{eq:pohybovka})).
The energy loss of a non-thermal electron $\Delta E$ with
kinetic energy $E$ and velocity $\varv$ caused by Coulomb collisions
in a partly ionised hydrogen cold target, per time-step $\Delta t$,
can be approximated by
   \begin{equation}
       \Delta E = -\frac{2\pi e^4}{E}\left[\Lambda x +
         \Lambda'(1-x)\right]n\, \varv\, \Delta t \ ,
     \label{eq:E_loss_Bai}
   \end{equation}
where $n = n_{\mathrm{p}} + n_{\mathrm{n}}$ is the number density of
equivalent hydrogen atoms, $n_{\mathrm{p}}$ and $n_{\mathrm{n}}$
are the proton and hydrogen number densities, respectively,
$x= n_{\mathrm{p}}/n$ is the hydrogen ionisation, and
$\Lambda$, $\Lambda'$ are the Coulomb logarithms \citep{emslie1978}.

The scattering due to Coulomb collisions is simulated using the Monte
Carlo method. According to \citet{bai1982}, the relation between
the rms of the scattering angle $\Delta\vartheta_{\mathrm{C}}$, the
ratio $\Delta E/E$, and the Lorentz factor $\gamma_{\mathrm{L}}$ is
\begin{equation}
       \overline{\Delta\vartheta_{\mathrm{C}}^2} = \left(\frac{\Delta E}{E}\right) \left(\frac{4}{\gamma_{\mathrm{L}}+1}\right)
     \label{eq:theta_scatt_Bai} \ ,
   \end{equation}
when $\overline{\Delta\vartheta_{\mathrm{C}}^2} \ll 1$ (or
equivalently, $\Delta E/E \ll 1$).
The value of scattering angle $\Delta\vartheta_{\mathrm{C}}$ is given by
a Gaussian distribution, the rms of which is computed by the equation (\ref{eq:theta_scatt_Bai}).

The change in the pitch angle caused by the magnetic force
$\vec{F}_{\mathrm{m}}$, see equation~(\ref{eq:pohybovka}), in the
region of magnetic field convergence is
\begin{equation}
\Delta\vartheta_{\mathrm{B}} =  \frac{B_{i+1}-B_i}{2B_i}\tan{\vartheta_i}
\ ,
\end{equation}
providing $(\Delta\vartheta_{\mathrm{B}})^2\ll1$ in a single time-step,
where $B_i$ and $B_{i+1}$ are the magnetic field strengths at the beginning
and end of the particle path, and $\vartheta_i$ is the initial pitch angle.
The total change of the pitch angle in a single time-step due to
  collisions and magnetic field non-uniformity is $\Delta\vartheta =
\Delta\vartheta_{\mathrm{C}}+\Delta\vartheta_{\mathrm{B}}$,
 and the new electron pitch angle $\vartheta$ is then
  obtained using the cosine rule from the spherical trigonometry
\begin{equation}
\cos\vartheta = \cos\vartheta_i \cos\Delta\vartheta + \sin\vartheta_i
\sin\Delta\vartheta \cos\varphi \ ,
\end{equation}
where $\varphi$ is the azimuthal angle given by a
uniform distribution $0\le\varphi<2\pi$.
More details concerning computer implementation can be found in
\citet{varady2005, varady2010} and \citet{kasparova2009}.

\subsection{Secondary accelerating mechanisms}

To include the secondary acceleration mechanisms, we added
either the static or stochastic electric fields that re-accelerate or
decelerate the test-particles with respect to the mutual directions
of the electric field and instantaneous test-particle velocities.
The interaction of the non-thermal particles with the
  re-accelerating electric field, the $-e\vec{E}$ term in equation~(\ref{eq:pohybovka}),
is calculated using the Boris relativistic
  algorithm \citep[see][Sect.~8.5.2]{perratt1992}.
The effects of the return current are not considered. Relatively low
electron fluxes transported from the corona
($\mathcal{F}_0/2 = 2.5\times10^9$~erg\,cm$^{-2}$\,s$^{-1}$
towards each footpoint) partially justify this negligence.

\begin{figure}[!t]
  \centerline{\includegraphics[width=8.2cm]{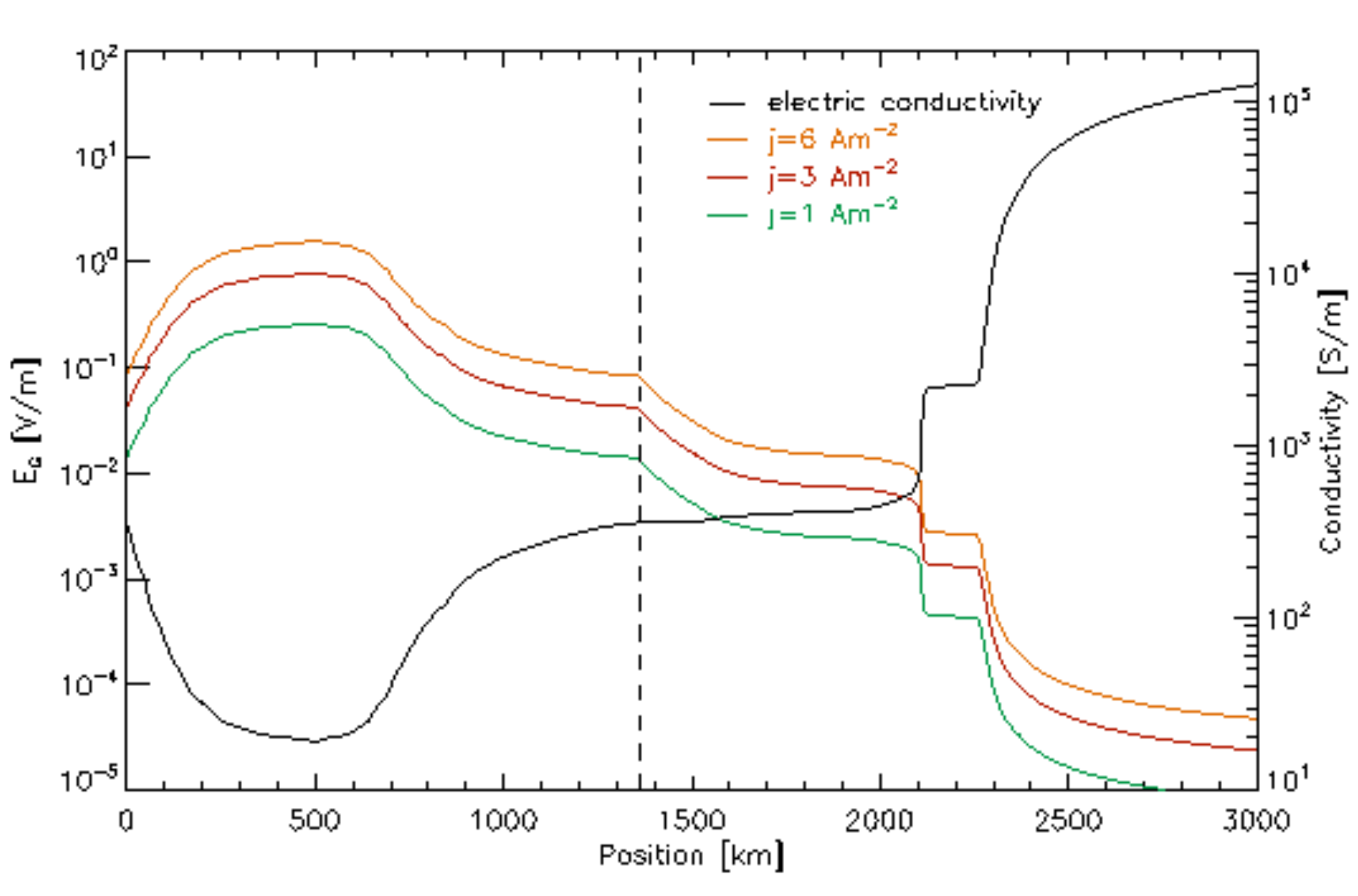}}
\caption{Classical electric conductivity $\sigma$ in the lower
  VAL~C atmosphere according to \citet{kubat1986} and the
  magnitude of the corresponding ${\bf E}_{\mathrm{G}}$ for
   various current densities
  ${\bf j}$.}
\label{fig:2}
\end{figure}

\subsubsection{Static electric field -- GRTTM}

We now consider a situation where electric currents flow in the flare
loop before and during the flare impulsive phase due to the non-zero
helicity of the pre-flare magnetic field \citep{karlicky1995}.
Furthermore, at the very beginning of the flare, the current-carrying loops are
unstable to the kink and tearing-mode instabilities, which produce
filamented electric currents in a natural way \citep{kuijpers1981, karlicky2010,
kliem2010, gordovskyy2011}. If electrons are accelerated in the coronal part of
the individual current thread, they propagate along it and interact with the
corresponding global re-acceleration resistive static electric field ${\bf E}_{\mathrm{G}}$
driving the current. The field corresponding to the current density ${\bf j}$ is
\begin{equation}
    {\bf E}_{\mathrm{G}}={\bf j}/\sigma \ ,
\end{equation}
where $\sigma$ is the plasma electric conductivity. The general
formula for plasma conductivity is
\begin{equation}
    \sigma = \frac{\omega_{\mathrm{pe}}^2}{4\pi \nu_{\mathrm{e}}}\ ,
\end{equation}
where $\omega^2_{\mathrm{pe}} = 4\pi  e^2 n_{\mathrm{e}} / m_{\mathrm{e}}$ is the
electron plasma frequency, and $\nu_{\mathrm{e}}$ the electron collisional
rate. In case electric currents propagate in plasma free of any plasma
waves, the collisional frequency corresponds to the classical value
\begin{equation}
    \nu_{\mathrm{e}} =
    2.91\times{10^{-6}}\frac{n_{\mathrm{e}}}{T_{\mathrm{e}}^{3/2}} \ {\Lambda}\ ,
\end{equation}
in the SI units, where $T_{\mathrm{e}}$ is the electron temperature.
On the other hand, the presence of plasma waves can increase the
collisional frequency to anomalous values: for the anomalous resistivity see
\citet{hey1981}.

To assess the influence of static electric field on the outcome of the
chromospheric bombardment by non-thermal electrons, we assume
a single thread of constant current density with magnitude below
any current instability thresholds. Then we calculate the magnitude of
corresponding direct field
${\bf E}_{\mathrm{G}}$ along the thread using the classical isotropic electric
conductivity obtained by \citet{kubat1986}. The  conductivity was
calculated using the updated values of proton--hydrogen scattering
cross-section for the quiet VAL C atmosphere (see Fig.~\ref{fig:2}). Owing to
temperature dependence of $\sigma$ and the convergence of magnetic field in the
chromosphere,  contributing to the increase in the local current density, the
resulting ${\bf E}_{\mathrm{G}}$ grows rather quickly in the chromosphere (see
Fig.~\ref{fig:2}). Furthermore,  ${\bf E}_{\mathrm{G}}$ tends to accelerate the
beam electrons towards one footpoint and to decelerate them towards the second
one, providing an asymmetric flare heating of the individual thread
footpoints. From now on, we refer to the individual footpoints as the
primary and the secondary footpoints, respectively and to this model as the
global re-accelerating thick-target model (GRTTM).

The steep increase in ${\bf E}_{\mathrm{G}}$, hence the high efficiency of
GRTTM, is essentially linked with the decrease in temperature in the
chromosphere. In contrast, we have already pointed out that chromospheric
plasma in flares is heated to temperatures up to  $10^5$~K on the timescales
$\le1$~s. Such an extreme increase in temperature substantially increases the
classical electric conductivity ($\sigma\propto T_{\mathrm{e}}^{3/2}$) in the
corresponding region, and by the same factor it decreases the electric field
${\bf E}_{\mathrm{G}}$, so the flare heating of the chromosphere should
basically cease the re-acceleration in the thick-target region very early after
the start of the impulsive phase. On the other hand, under the flare
conditions, generation of a high anomalous resistivity could be expected due to
plasma instabilities, so the accelerating mechanism could continue working.

\subsubsection{Stochastic electric fields -- LRTTM}

\begin{figure}[!t]
  \centerline{\includegraphics[width=8.2cm]{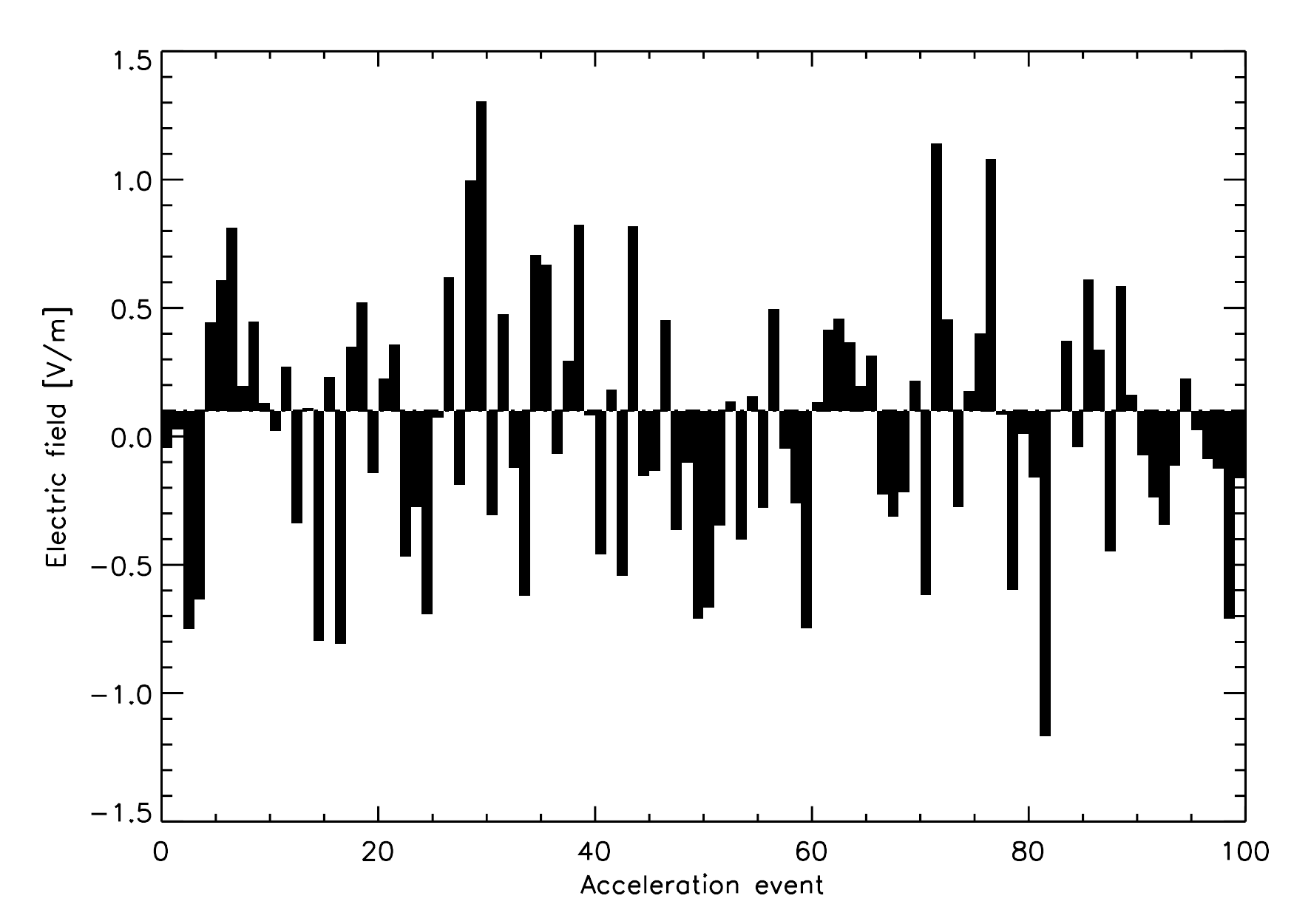}}
  \centerline{\includegraphics[width=8.2cm]{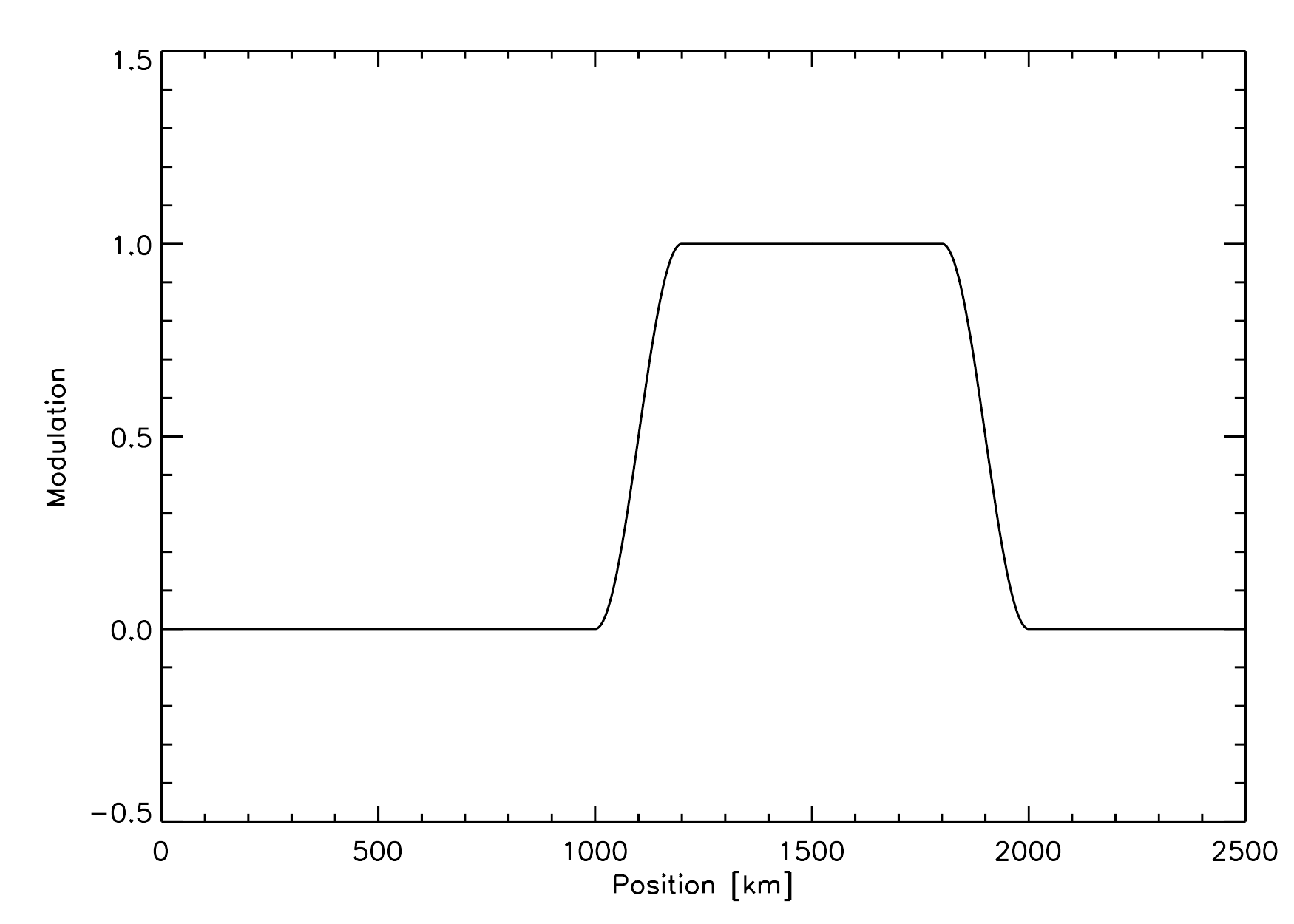}}
\caption{{\it Top:} Example of ${{\bf E}_{\mathrm{L}}}$-II
type stochastic electric field with $\overline{E_{\mathrm{L}}} = 0.1$~V\,m$^{-1}$ and
  $\mbox{var}(E_{\mathrm{L}})=0.5$~V\,m$^{-1}$ corresponding to the
 distribution function in  Fig.~\ref{fig:13}.
 {\it Bottom:} The the spatial modulation of ${\bf E}_{\mathrm{L}}$.}
 \label{fig:3}
\end{figure}

Inspired by \citet{brown2009} and \citet{turkmani2012},
we produced a simplified local re-acceleration thick-target model
(LRTTM). To approximate the distribution of electric fields arising as
a consequence of a current sheet cascade in the randomly stressed
magnetic fields \citep{turkmani2005, turkmani2006}, we assume a
region (between 1~--~2~Mm) of stochastic re-acceleration electric field
${{\bf E}_{\mathrm{L}}}$, spatially modulated by the function shown in
Fig.~\ref{fig:3} (bottom). The position of the local
re-acceleration region is one of the free parameters of the model.
It roughly corresponds to the chromosphere and encompasses the
regions of magnetic field convergence and the rapid change of hydrogen
ionisation (see Fig.~\ref{fig:1}).

The stochastic electric fields ${{\bf E}_{\mathrm{L}}}$ are
generated only in the directions parallel and anti-parallel relative
to the loop axis, and their distribution corresponds to Gaussians with
various mean values $\overline{E_{\mathrm{L}}}$ and variances
$\mbox{var}(E_{\mathrm{L}}) = \overline{E_{\mathrm{L}}^2} -\overline{E_{\mathrm{L}}}^2.$
We examine two types of  ${{\bf E}_{\mathrm{L}}}$:
\begin{enumerate}[{${{\bf E}_{\mathrm{L}}}$}-I.]
\item{A stochastic electric field with zero mean value}
\begin{equation}
\overline{E_{\mathrm{L}}}=0 \ , \ \ \ \ \ \
\mbox{var}({E_{\mathrm{L}}})>0 \ .
\end{equation}
\item{A combination of spatially localised static electric field
with a stochastic component  (see Fig.~\ref{fig:3})}
\begin{equation}
\overline{E_{\mathrm{L}}}\ne0 \ , \ \ \ \ \
\mbox{var}{(E_{\mathrm{L}}})\ge0 \ .
\end{equation}
\end{enumerate}
In the case of ${{\bf E}_{\mathrm{L}}}$-II, the sign of the static component
$\overline{E_{\mathrm{L}}}$ always assures acceleration of the
non-thermal electrons towards the nearest footpoint.
This field type can develop in the thick-target region if
stochastic fields are present in a globally twisted magnetic loop.
In comparison with the GRTTM, the LRTTM is characterised by
abrupt changes in magnitude and orientation of the accelerating or
decelerating electric fields representing the individual current
sheets in the thick-target region \citep{turkmani2006} (compare
Figs.~\ref{fig:2} and \ref{fig:3}).

The integration of motion of individual beam electron clusters
for the LRTTM is performed in the following way.
 In each time-step (corresponding to $\Delta t=5\times10^{-5}$~s),
we generate a random value of ${\bf E}_{\mathrm{L}}$ for each particle
within the acceleration region. In this way we model
the situation where the beam electrons are moving
in the stochastic electric fields, whose configuration temporally
changes. Therefore, the electrons only have a
negligible chance of passing through exactly the same configuration of current
sheets and of experiencing the same acceleration (deceleration) sequence.
The  time-step basically determines the spatial extent of the
  individual current sheets. In order to keep the size independent of
  particle velocities, we weight ${\bf E}_{\mathrm{L}}$ using a factor
  $\varv_0/\varv$, where $\varv_0$ and $\varv$ are the velocities corresponding to
  the low-energy cutoff and to the particular particle, respectively. The
  time-step $\Delta t=5\times10^{-5}$~s thus corresponds to the current
  sheet size $\sim$3~km. Simulations with various time-steps showed
  that the results are not very sensitive to the choice of the time-step.
Using the weighted value of ${\bf E}_{\mathrm{L}}$ we relativistically move
the electron from the old to the new position. Then we calculate the
energy loss and scattering due to the passage of the particle through
the corresponding column of plasma and the effects of converging
magnetic field. This is done repeatedly for the
whole population of test-particles. The corresponding total energy
deposit and  HXR spectrum are then calculated.

\subsection{HXR spectra}

The intensity $I(\epsilon, s)$
[photons\,cm$^{-2}$\,s$^{-1}$\,keV$^{-1}$] of HXR
bremsstrahlung observed on energy $\epsilon$,
emitted by plasma at a position $s$ along the flare loop, detected in
the vicinity of the Earth, was calculated using the
formula \citep{brown1971}
\begin{equation}
I(\epsilon, s) = \frac{n_{\mathrm{p}}(s) V(s)}{4\pi R^2}
\int_\epsilon^\infty  Q(E,\epsilon)\; \varv(E)\; n(E, s) \,\mathrm{d} E  \ .
\end{equation}
Here, $n_{\mathrm{p}}(s) V(s)$ is the total number of protons in the emitting
plasma volume $V(s)$ at a position $s$, distance $R=1$~AU,
$\varv(E)$ is the electron velocity calculated relativistically
from the electron energy, and $n(E, s)$ is
the number density of non-thermal electrons per energy in the emitting volume having
kinetic energy $E$.  The cross section $Q(E,\epsilon)$ for
bremsstrahlung was calculated using a semi-relativistic formula
given by \citep{haug1997}, multiplied by the Elwert factor
\citep{elwert1939}, considering the limit case when the entire
electron kinetic energy is emitted. The precision of the method
should be better than 1~\% for energies $\le300$~keV \citep{haug1997}.
To calculate the emitting volume $V(s)$ we assume a circular cross
section of the converging loop with a radius $r(s)=1.5\sqrt{B_{\mathrm{0}}/B(s)}$~Mm.
The HXR spectra are calculated on a spatial (height) grid $(s, s+\Delta s)$.
The individual emitting volumes along the grid are then $V(s)= \pi r(s)^2 \Delta s$.

\begin{figure*}[t]
  \centerline{
   \includegraphics[width=7.8cm]{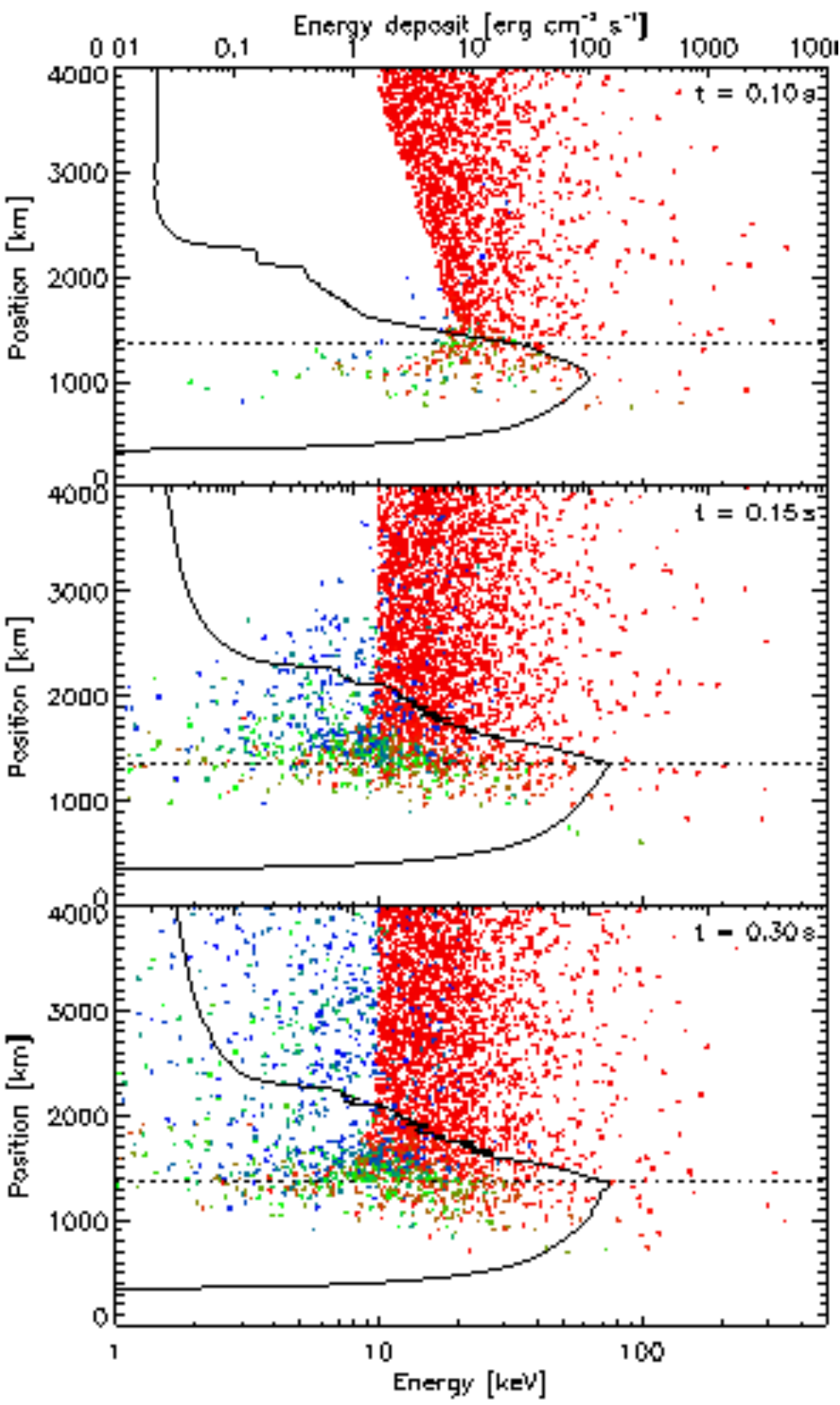}
   \includegraphics[width=7.8cm]{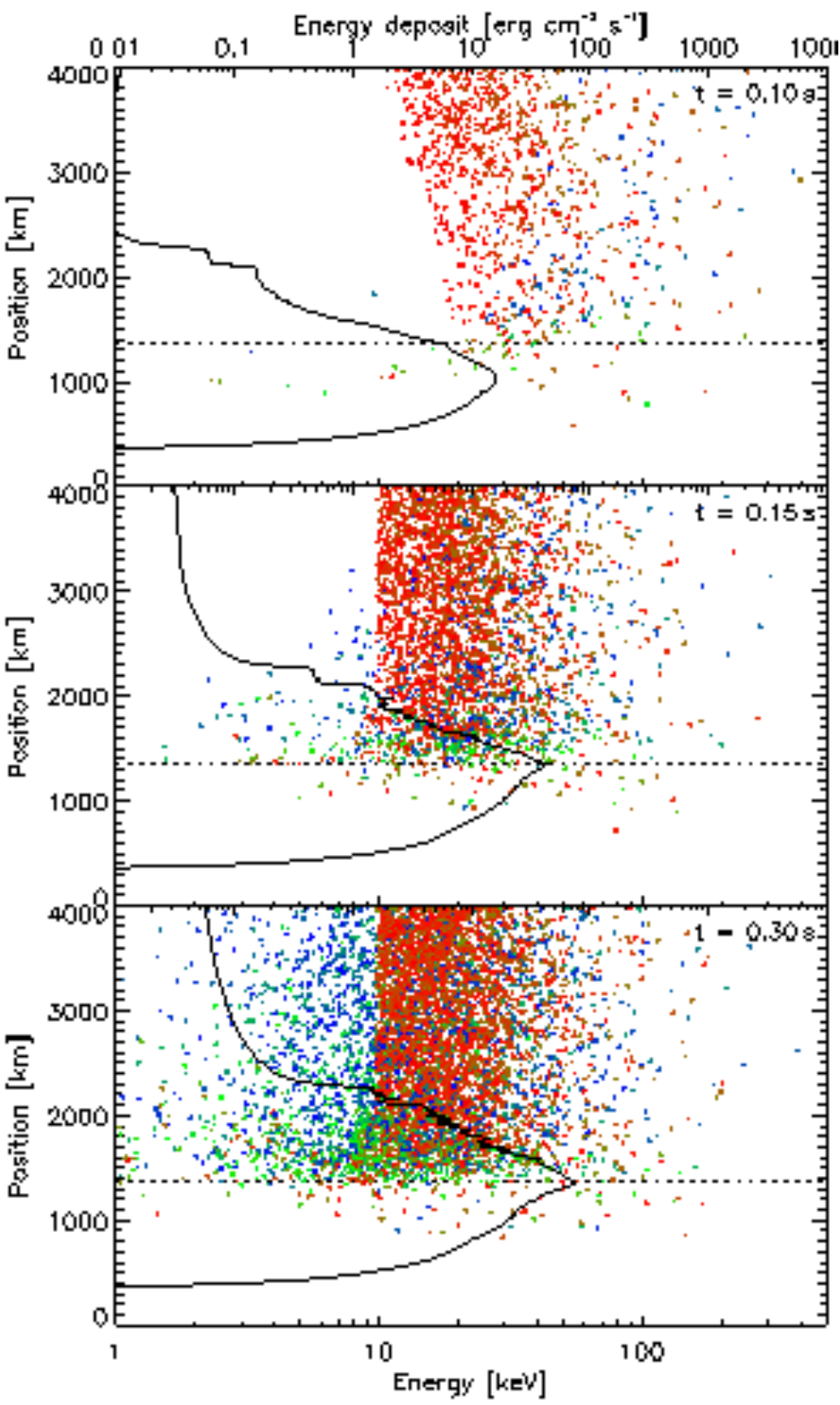}}
   \vspace{0em}
   \centerline{\includegraphics[width=6cm]{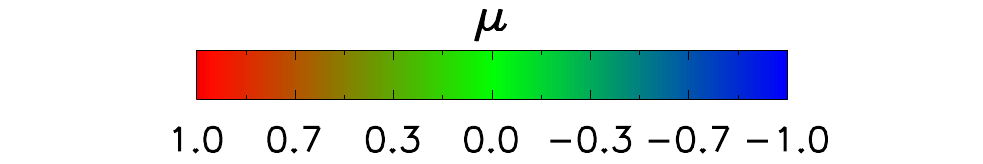}}
\caption{CTTM time evolution of distribution functions of non-thermal
electron energies versus positions with a colour coded $M(\mu_0)$ in
the VAL~C atmosphere. {\it Left:} $M^\mathrm{FF}$, {\it right:}
$M^\mathrm{SU}$. {\it From top to bottom:} individual snapshots at
$t= 0.1, 0.15, 0.3$~s after the beam injection into the loop at its
apex. The solid lines indicate the instantaneous energy deposits
corresponding to
${\mathcal{F}}_{\mathrm{0}}/2= 2.5\times10^{9}$~erg\,cm$^{-2}$\,s$^{-1}$.
The dotted horizontal lines indicate the bottom boundary of the magnetic mirror.
Only the lower part of the loop and one footpoint are displayed.}
\label{fig:4}
\end{figure*}

\section{Results}\label{sec:results}

We now concentrate on a comparison of outcomes of chromospheric bombardment for two
modifications of CTTM with the CTTM itself. In this section we present the
non-thermal electron distribution functions in the vicinity of footpoints and
several properties of the corresponding energy deposits and HXR intensities and
spectra. The quantitative results for the CTTM, GRTTM, and both considered
  types of LRTTM are summarised in Figs.~\ref{fig:GRTTM_tab}, \ref{fig:LRTTM_a_new}, and
   \ref{fig:LRTTM_b_new} and Tables~\ref{tab:CTTM}, \ref{tab:GRTTM},
   and \ref{tab:LRTTM}, respectively (Tables~\ref{tab:GRTTM} and
   \ref{tab:LRTTM} available only in the online version). Here,
the factor $\mathcal{F_{\mathrm{R}}}/\mathcal{F}_{\mathrm{0}}$ gives the ratio
of the reflected (due to the magnetic mirroring, re-acceleration, and
backscattering) to the original non-thermal electron energy flux coming from
the corona at position $s= 3$~Mm, measured at $t=0.3$~s after the beam
injection into the loop at its apex. To assess the magnitude of the energy
deposits for the individual models, we calculate the total energy deposited into
the chromosphere along a magnetic flux tube as
\begin{equation}
\mathcal{E}_{\mathrm{ch}}= \int\limits_{\mathrm{chromosphere}}
{E_{\mathrm{dep}}(s)} \mathrm{d}{V(s)} = S_0 B_0
\int\limits_{0}^{2.3\mathrm{Mm}} \frac{E_{\mathrm{dep}}(s)}{B(s)}
\mathrm{d}{s} \
\label{eq:15}
\end{equation}
and give the position of the energy deposit maximum
$s_{\max}$ in the atmosphere.
The factor $B_0/B(s)$ in integral (\ref{eq:15}) accounts for the
convergence of the magnetic field, $E_{\mathrm{dep}}(s)$ is the local
energy deposit in units [erg\,cm$^{-3}$\,s$^{-1}$], and the limits of
integration correspond to the upper and lower
boundaries of the chromosphere. The lower limit lies far below the
stopping depths of the beam electrons for all the studied models.
When all the beam energy is deposited into the chromosphere and
$S_0=1$~cm$^2$, the value of $\mathcal{E}_{\mathrm{ch}}$ in units
[erg\,s$^{-1}$] corresponds to the value of the initial flux
$\mathcal{F}_{\mathrm{0}}$.

For HXR we give the intensity $I_{25\,\mathrm{keV}}$ and the power-law index
$\gamma_{25\,\mathrm{keV}}$  measured at energy 25~keV. Furthermore,
we applied the RHESSI spectral analysis
software\footnote{\url{http://hesperia.gsfc.nasa.gov/rhessi2/home/software/spectroscopy/spectral-analysis-software/}}
(OSPEX) to modelled total X-ray spectra to imitate common spectral
analysis. We assumed that
these spectra were incident on RHESSI detectors and forward-fitted the
``detected'' count spectra. In the fitting we used the OSPEX
  thick-target model and a single power-law injected electron spectrum.
In this way we obtained  the fitted electron beam parameters.
To account for the non-uniform ionisation structure of the X-ray emitting
atmosphere, the fitting function \texttt{f\_thick\_nui} in the step-function
mode was chosen. When the fitted parameters of \texttt{f\_thick\_nui}
were unrealistic and the X-ray emission was formed deep in the layers of almost
neutral plasma, \texttt{f\_thick} with neutral energy loss term was used. Also,
we modified the standard OSPEX energy loss term and the ratio of Coulomb
logarithms to be consistent with relations used in the test-particle code.
The results of this analysis, the fitted energy flux $\mathcal{F}_{\mathrm{0}}'$,
the power-law index $\delta_\mathrm{p}'$, and the low-energy cutoff $E_0'$
are listed in Tables\footnote{Tables~2 and 3 are online only.} \ref{tab:CTTM},
\ref{tab:GRTTM}, and \ref{tab:LRTTM} and displayed in Figs.~\ref{fig:GRTTM_tab},
\ref{fig:LRTTM_a_new}, and \ref{fig:LRTTM_b_new}.

\begin{figure*}[!t]
\centerline{\includegraphics[width=8.cm]{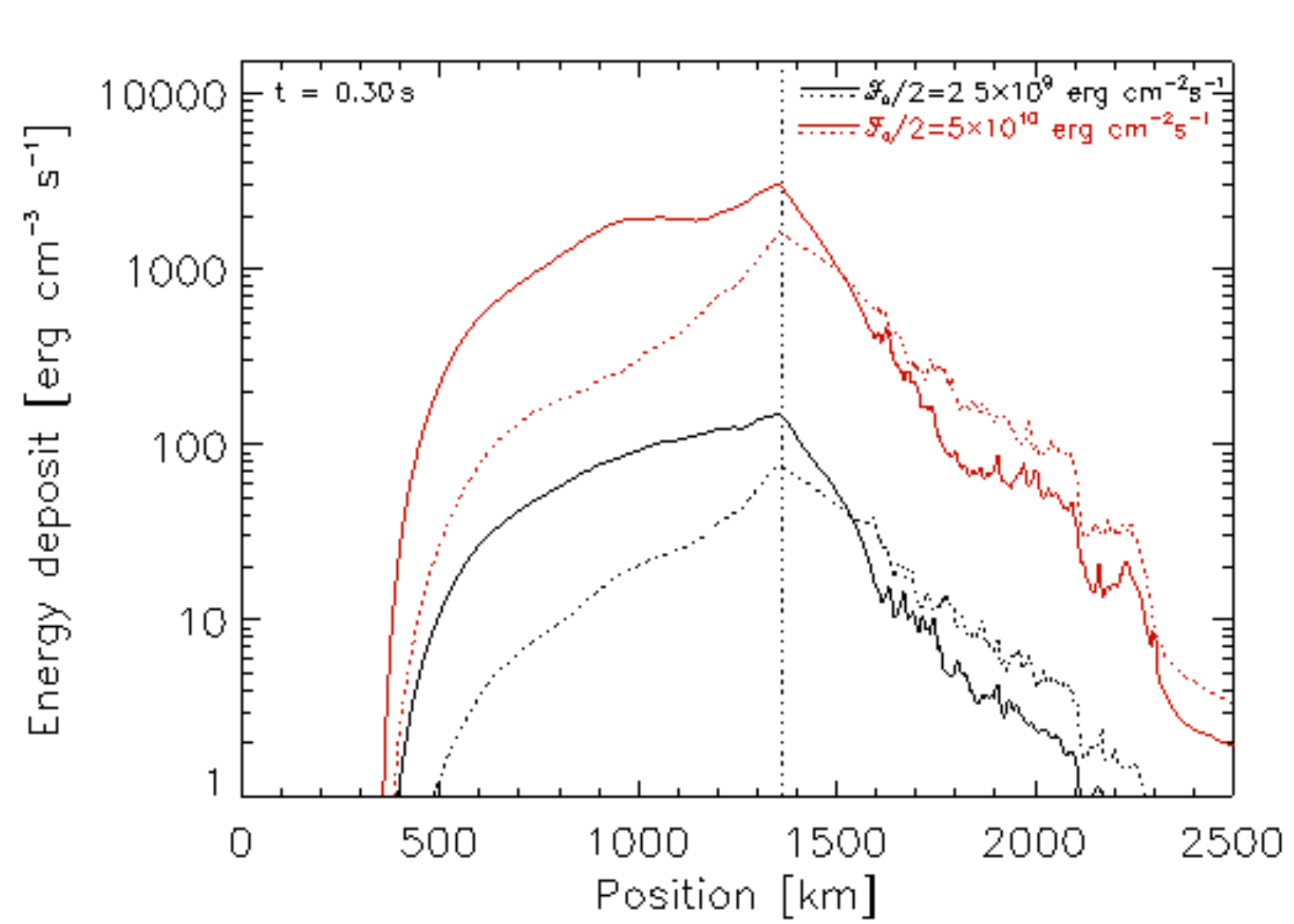}
              \includegraphics[width=8.cm]{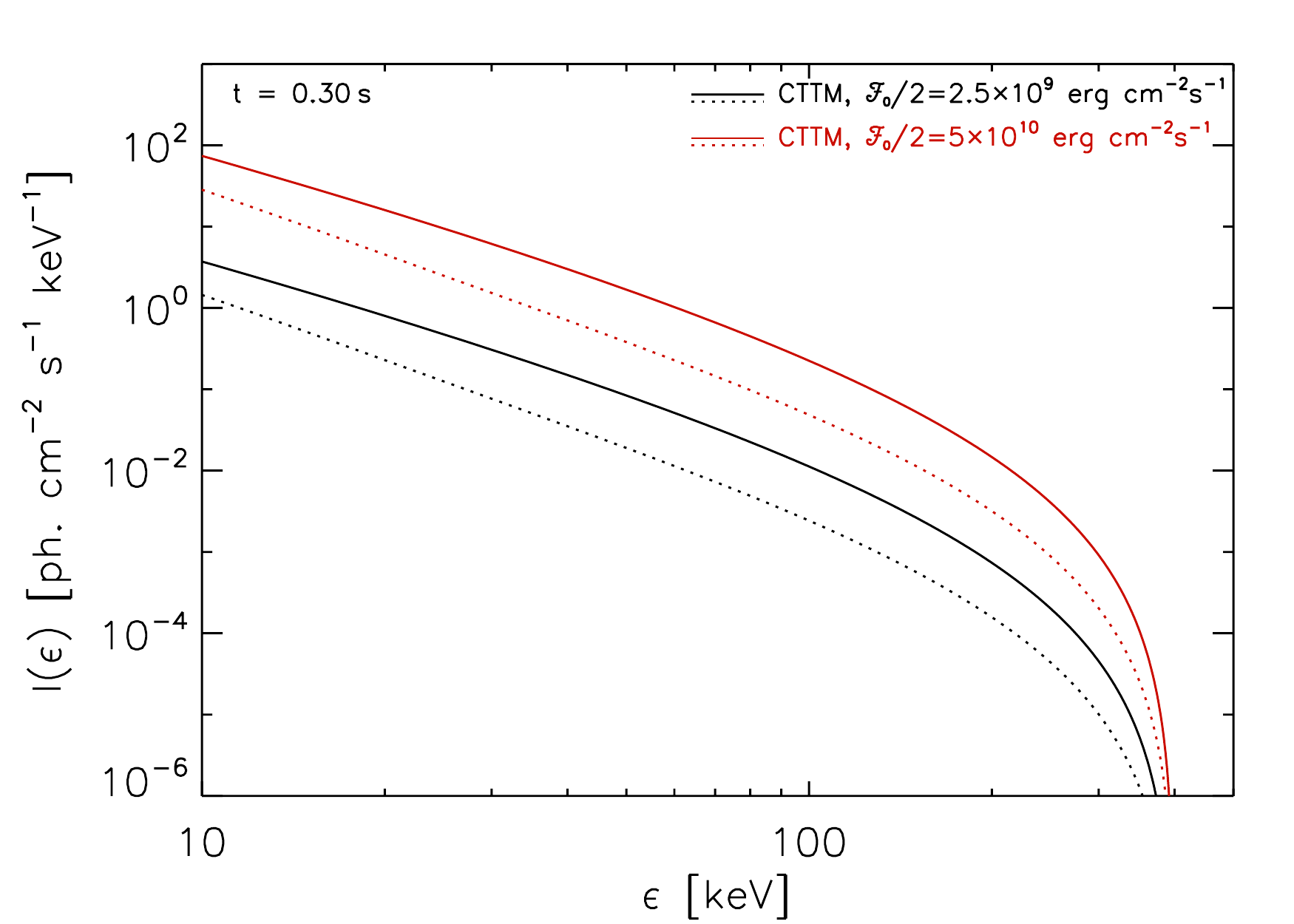}}
\caption{{\it Left:} CTTM instantaneous energy deposits into the
  VAL~C  atmosphere at $t$=0.3~s for energy fluxes $\mathcal{F}_{\mathrm{0}}/2
  =5\times10^{10}$~erg\,cm$^{-2}$\,s$^{-1}$ (red lines) and $\mathcal{F}_{\mathrm{0}}/2
   =2.5\times10^9$~erg\,cm$^{-2}$\,s$^{-1}$ (black lines).
  The dotted vertical line indicates the bottom boundary of the
  magnetic mirror. {\it Right:} The HXR spectra integrated over one
  half of the loop. In both panels the solid
   lines represent $M^{\mathrm{FF}}$, the dotted lines
   $M^{\mathrm{SU}}$ case. }
\label{fig:5}
\end{figure*}

\subsection{CTTM}

To produce a basis for comparison we present results for the classical
CTTM in a converging magnetic field. The information on kinematics of
non-thermal electrons for both initial $\mu$-distributions we
considered is
incorporated into Fig.~\ref{fig:4}. We first concentrate on
the left-hands panels showing the time dependent distributions for
$M^{\mathrm{FF}}$ case. The top panel for $t=0.1$~s corresponds
to the transition state when the loop is being filled with  non-thermal
electrons. The process of filling is apparent as a depletion of the
distribution function at low energies in the region ranging from approximately
$1.4$~Mm to $3.7$~Mm.  The distribution above the low-energy
  cutoff and the bottom boundary of the magnetic mirror is dominated by
red, so a vast majority of particles move downwards with
$\mu\approx1$. At low energies ($E<20$~keV), a low-energy tail of
particles starts to form in the region under the lower boundary of
the magnetic mirror.
It consists of particles with originally higher energies
that lost part of their energy owing to their interactions with the
target plasma. The tail is rich in particles with $\mu\approx0$
(green), and it also contains a few back-scattered particles
with $\mu\approx-1$ (blue).
Coulomb scattering leads to an increase in pitch angles of low-energy electrons in
the region above the magnetic mirror. These particles do not satisfy the condition for passing
through the mirror. They
are reflected and propagate back to the loop top and fill the loop with
a population of low-energy electrons ($<20$~keV) with $-1\le\mu<0$.

Such a low-energy tail is more clearly pronounced in the
subsequent times in the vicinity and slightly above the lower boundary
of the magnetic mirror.
The following snapshot for $t = 0.15$~s, when even the particles with
lowest energies reached the thick-target region, shows the proceeding
thermalisation of  beam electrons in this region and increase in
particle number with $\mu\le0$ in the low-energy tail.
A new population of particles  with $\mu\approx-1$ starts to form
and propagate upwards, towards the loop top.
The snapshot at $t=0.3$~s roughly corresponds to a fully developed
state. The part of the distribution function at the vicinity of the
bottom boundary of the magnetic mirror and in the low-energy region
$E<20$~keV is dominated by particles with $\mu\approx0$.
The reflected energy flux propagating upwards is approximately
4\% of the original flux $\mathcal{F}_{\mathrm{0}}$  for the
$M^\mathrm{FF}$ case (see Table~\ref{tab:CTTM}).

The distribution functions corresponding to $M^{\mathrm{SU}}$ are shown in
Fig.~\ref{fig:4} (right). The overall behaviour of the beam electrons
is quite similar to the previously discussed case. The most obvious difference
is the enhancement of the particle populations with $\mu<0$ on all energies
(corresponding to 40\% of the initial flux $\mathcal{F}_{\mathrm{0}}$) and
$\mu\approx0$ predominantly on low energies ($E<40$~keV) localised above the
bottom boundary of the magnetic mirror. The differences between the
$M^{\mathrm{FF}}$ and $M^{\mathrm{SU}}$ cases naturally influence the resulting
energy deposits and properties of the corresponding HXR emission (see
 Figs.~\ref{fig:4}, \ref{fig:5}).
The CTTM in the adopted arrangement gives identical results
for both footpoints. Therefore for $t>0.3$~s the particles
reflected at the second footpoint reach the loop top and appear as
a new population of particles moving downwards to the first footpoint.
For simplicity we only concentrate on times $t\le0.3$~s.

\begin{table*}
\caption{Summary of results for the CTTM.} 
\label{tab:CTTM} 
\begin{center}
\begin{tabular}{c c c c c c c c c} 
\hline\hline 
$\mathcal{F}_0/2\times10^{9}$ &
$\mathcal{F_{\mathrm{R}}}/\mathcal{F}_{\mathrm{0}}$ &
$\mathcal{E}_{\mathrm{ch}}/10^{9}$ & $s_{\max}$
& $I_{25\,\mathrm{keV}}$ &$\gamma_{25\,\mathrm{keV}}$ &
$\mathcal{F}_{\mathrm{0}}'/2\times10^{9}$ & $\delta'_\mathrm{p}$ & $E'_0$ \\

[erg\,cm$^{-2}$\,s$^{-1}$] & [\%]&  [erg\,s$^{-1}$] & [Mm] &
[cm$^{-2}$\,s$^{-1}$\,keV$^{-1}$] &  & [erg\,cm$^{-2}$\,s$^{-1}$]& & [keV]  \\ 
\hline 
\multicolumn{7}{c}{$R_{\mathrm{m}}=1$} \\ \hline
2.5 & 0.08 (2.9) & 2.3 (2.2) & 1.3 (1.4) & 0.45 (0.44) & 2.4 (2.4) &
2.6 (2.4) & 3.0 (3.0) & 10 (11) \\
50  & 0.25 (3.0) & 47 (44) & 1.3 (1.4) & 9.0 (8.8) & 2.4 (2.4) & 52
(48) &  3.0 (3.0) & 10 (11) \\
\hline 
\multicolumn{7}{c}{$R_{\mathrm{m}}=5$} \\ \hline
2.5 & 4.3 (40) & 2.2 (1.4) & 1.4 (1.4) & 0.45 (0.12) & 2.4 (2.7) & 2.5
(1.3) & 3.0 (3.4) & 11 (10)\\
50  & 3.9 (40) & 46 (28) & 1.4 (1.4) & 9.1 (2.4) & 2.4 (2.7) & 53 (25)
&  3.0 (3.4) &10 (10) \\
\hline
\end{tabular}
\end{center}
\tablefoot{
$\mathcal{F}_{\mathrm{0}}$ -- the initial energy flux,
$\mathcal{F_{\mathrm{R}}}/\mathcal{F}_{\mathrm{0}}$ -- ratio
of reflected to initial energy flux at
$s= 3$\,Mm and $t=0.3$\,s,  $\mathcal{E}_{\mathrm{ch}}$ --
integrated chromospheric energy deposit, $s_{\mathrm{max}}$ --
position of energy deposit maximum, $I_{25\,\mathrm{keV}}$,
$\gamma_{25\,\mathrm{keV}}$ --  HXR intensity and power-law index
measured at energy 25\,keV, $\mathcal{F}_{\mathrm{0}}'$,
$\delta_\mathrm{p}'$, and $E_0'$ -- the fitted
values of energy flux, power-law index, and low-energy cutoff, respectively.
 The non-parenthetical and parenthetical values are for the
  $M^{\mathrm{FF}}$ and $M^{\mathrm{SU}}$ cases of  $M(\mu_0)$,
  respectively. Applies to further online tables.}
\end{table*}
\begin{figure*}[!t]
  \centerline{\includegraphics[width=8.8cm]{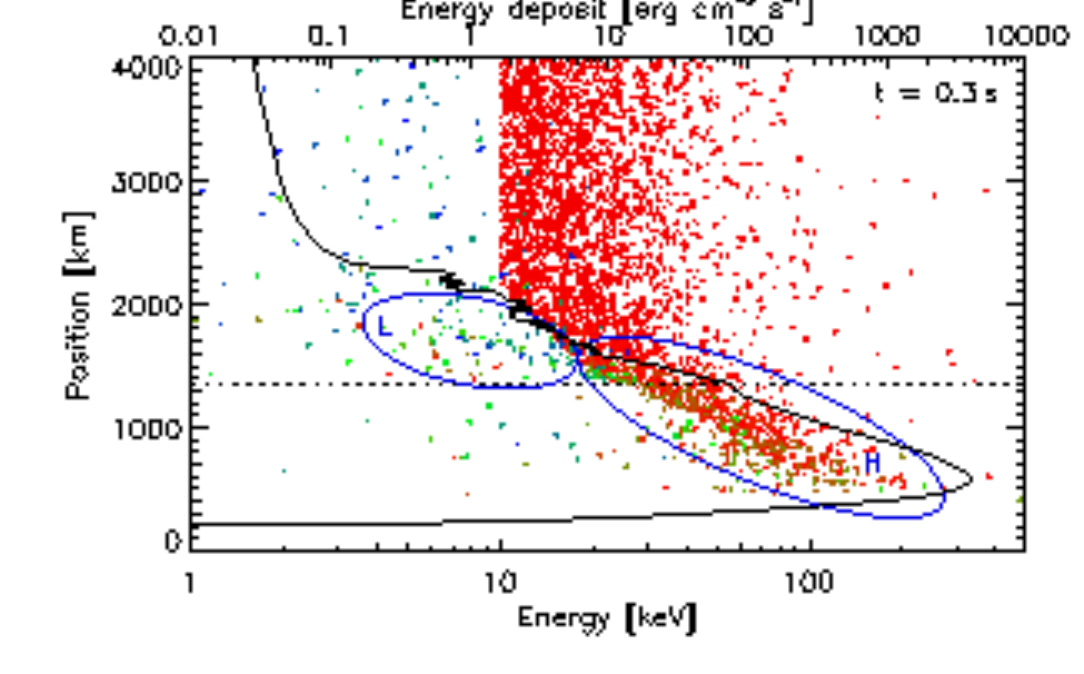}
              \includegraphics[width=8.8cm]{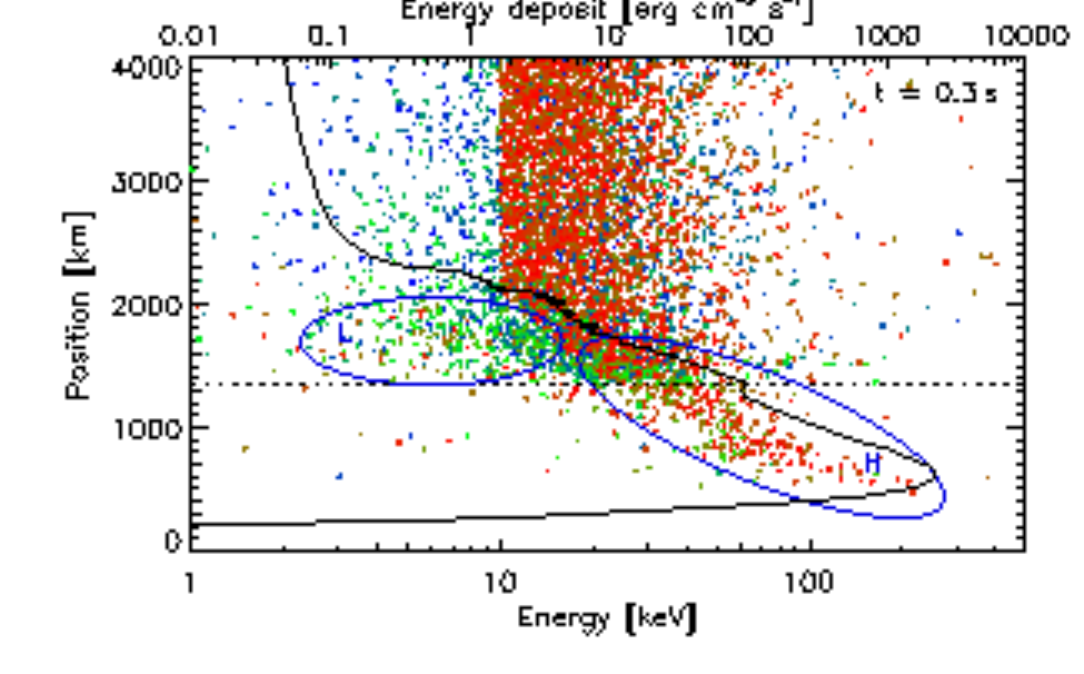}}
  \centerline{\includegraphics[width=8.8cm]{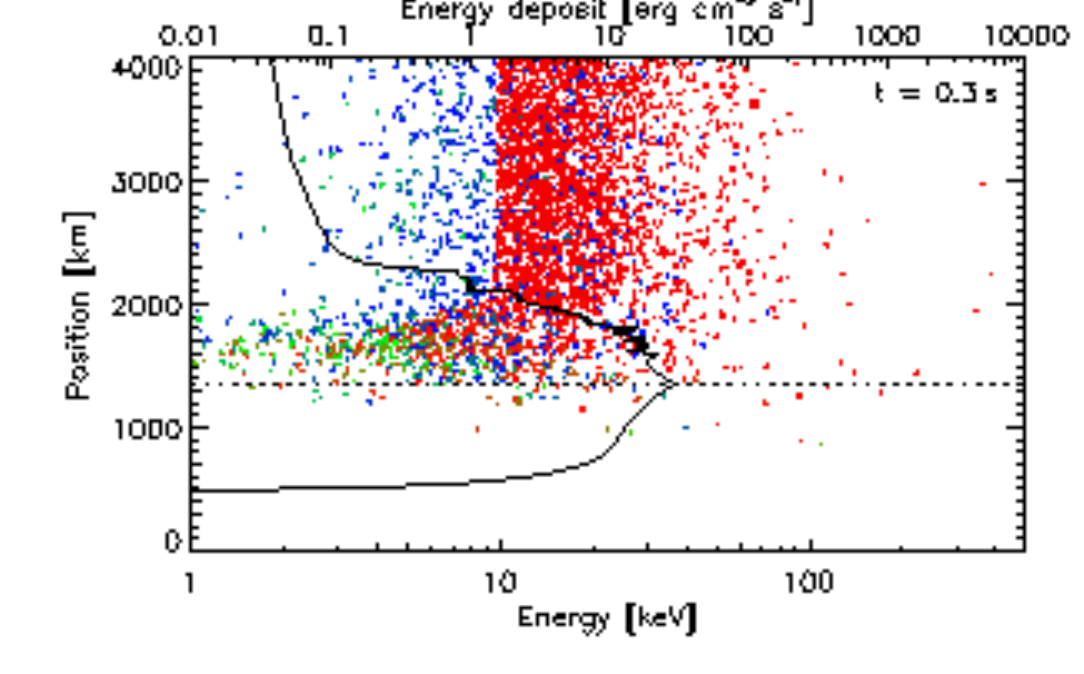}
              \includegraphics[width=8.8cm]{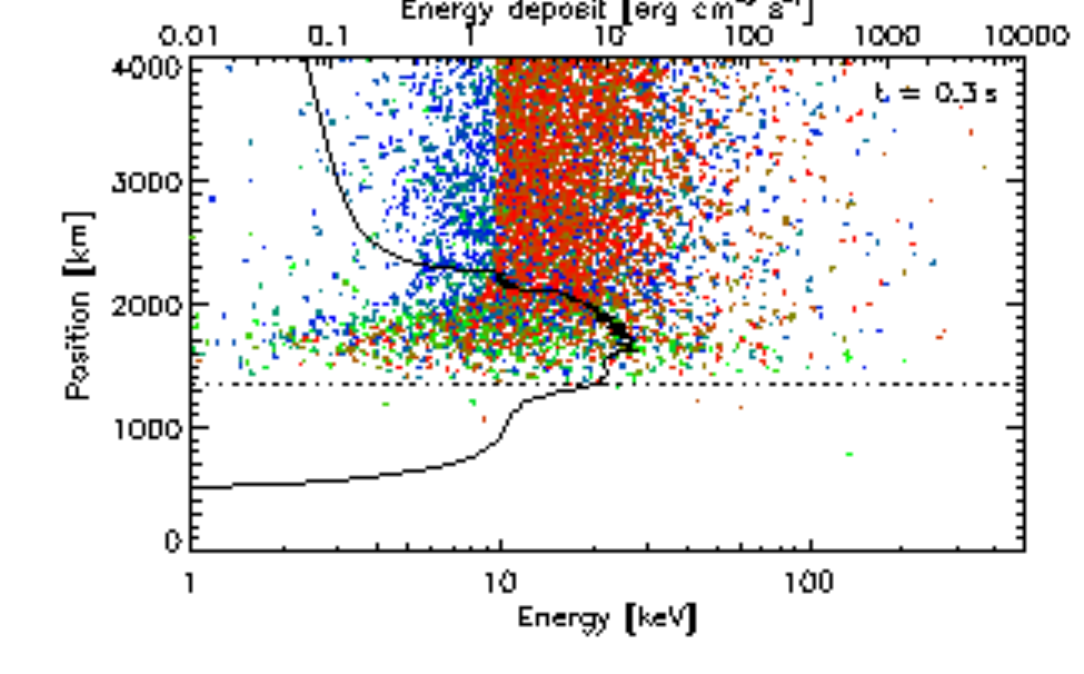}}
\vspace{-2em}
\centerline{\includegraphics[width=6cm]{fig_0.pdf}}
\caption{GRTTM distribution functions of non-thermal electron energies versus
  positions with a colour coded $M(\mu_0)$ corresponding
  to the current density $j=6$\,A\,m$^{-2}$ in the VAL C
  atmosphere at time $t=0.3$\,s after the beam injection into the
  loop at its apex.
{\it Top:} primary footpoint, {\it bottom:} secondary footpoint,
{\it left:} $M^{\mathrm{FF}}$,
{\it right:} $M^{\mathrm{SU}}$.
The solid lines indicate the instantaneous energy deposits
corresponding to $\mathcal{F}_{\mathrm{0}}/2=
2.5\times10^9$~erg\,cm$^{-2}$\,s$^{-1}$, the dotted horizontal lines
the bottom boundary of the magnetic mirror and the blue ellipses
labelled L and H denote tails in the particle distribution function.
Only the vicinity of the footpoints are displayed.}
\label{fig:6}
\end{figure*}

To distinguish the effects of the $\mu$-distribution and magnetic field
convergence, Table~\ref{tab:CTTM} also lists the characteristics of CTTM for the
case of no magnetic mirror, i.e. $R_\mathrm{m}=1$. It shows that it is the
magnetic field convergence that significantly influences
$\mathcal{F_{\mathrm{R}}}/\mathcal{F}_{\mathrm{0}}$ and
$\mathcal{E}_{\mathrm{ch}}$ in the case of $M^{\mathrm{SU}}$.

A comparison of energy deposits for both considered
initial $\mu$-distributions is shown in Fig.~\ref{fig:5} (left).
Because the adopted energy flux for
both models considering secondary re-acceleration
$\mathcal{F}_{\mathrm{0}}/2 = 2.5\times10^9$~erg\,cm$^{-2}$\,s$^{-1}$
is unrealistically low in the context of CTTM and flare physics, we also plot
energy deposits for the much higher and more realistic value
${\mathcal{F}}_{\mathrm{0}}/2 = 5\times10^{10}$~erg\,cm$^{-2}$\,s$^{-1}$. The results
corresponding to this flux will be used as a basis for comparison with
the energy deposits and HXR spectra obtained from the models involving the secondary
acceleration mechanisms.
The chromospheric energy deposit $\mathcal{E}_{\mathrm{ch}}$ scales
linearly with ${\mathcal{F}}_{\mathrm{0}}$ (see Table~\ref{tab:CTTM}), and
the positions of energy deposit maxima are almost
identical for all the considered cases approximately corresponding to the
placement of the lower boundary of the magnetic mirror
$s_{\max} = 1.36$\,Mm.  The peak in the energy
deposits at $s_{\max}$ and their steep decrease above it
(see Fig.~\ref{fig:5}, left) are caused by the constricted magnetic flux tube.
The influence of the initial $\mu$-distribution is obvious. For
the $M^{\mathrm{FF}}$ case, particles have a
greater chance of passing through the magnetic
mirror and thus of depositing their energy into the deeper layers. In the
$M^{\mathrm{SU}}$ case, when the particles reach the thick-target region and the
region of strongly converging field, their pitch angles are generally
higher: compare the left-hand and right-hand panels of Fig.~\ref{fig:4}.
Therefore the probability that an electron passes through the magnetic
mirror is strongly reduced. This naturally explains the systematic
enhancements in the energy deposits for $M^{\mathrm{SU}}$ in the layers above
and the decrease in the layers below the lower boundary of the magnetic
mirror in comparison with the $M^{\mathrm{FF}}$ case.

The corresponding HXR spectra are shown in Fig.~\ref{fig:5} (right),
and their parameters are summarised in Table~\ref{tab:CTTM}.
As expected, the HXR intensity $I_{25\,\mathrm{keV}}$ scales
linearly with the chromospheric deposit $\mathcal{E}_{\mathrm{ch}}$ or
the energy flux ${\mathcal{F}}_{\mathrm{0}}$.
Majority of the total X-ray emission, i.e. summed over the whole loop, comes
from the regions below the bottom boundary of the magnetic mirror. As
explained above, the number of particles passing through the magnetic
mirror is lower in the $M^{\mathrm{SU}}$  case than for
$M^{\mathrm{FF}}$, therefore the HXR emission corresponding to $M^{\mathrm{FF}}$
is more intense than the emission of $M^{\mathrm{SU}}$.

HXR spectra are steeper in the $M^{\mathrm{SU}}$ case owing to
presence of magnetic field convergence -- compare
$R_{\mathrm{m}}=1$ and 5 in Table~\ref{tab:CTTM}.
Fitted beam injected energy flux agrees well (within 20\%) with the
$\mathcal{E}_{\mathrm{ch}}$, whereas $\delta_\mathrm{p}'$ and $E_0'$ are the
same as those of the injected power law. An exception is the larger
$\delta_\mathrm{p}'$ in the $M^{\mathrm{SU}}$ case, which corresponds
to the mentioned HXR spectral behaviour and the fact that
the spectral fitting does not take the scattering induced
by change in $B$ into account.

\subsection{GRTTM}

\begin{figure*}
\centerline{\includegraphics[width=8.8cm]{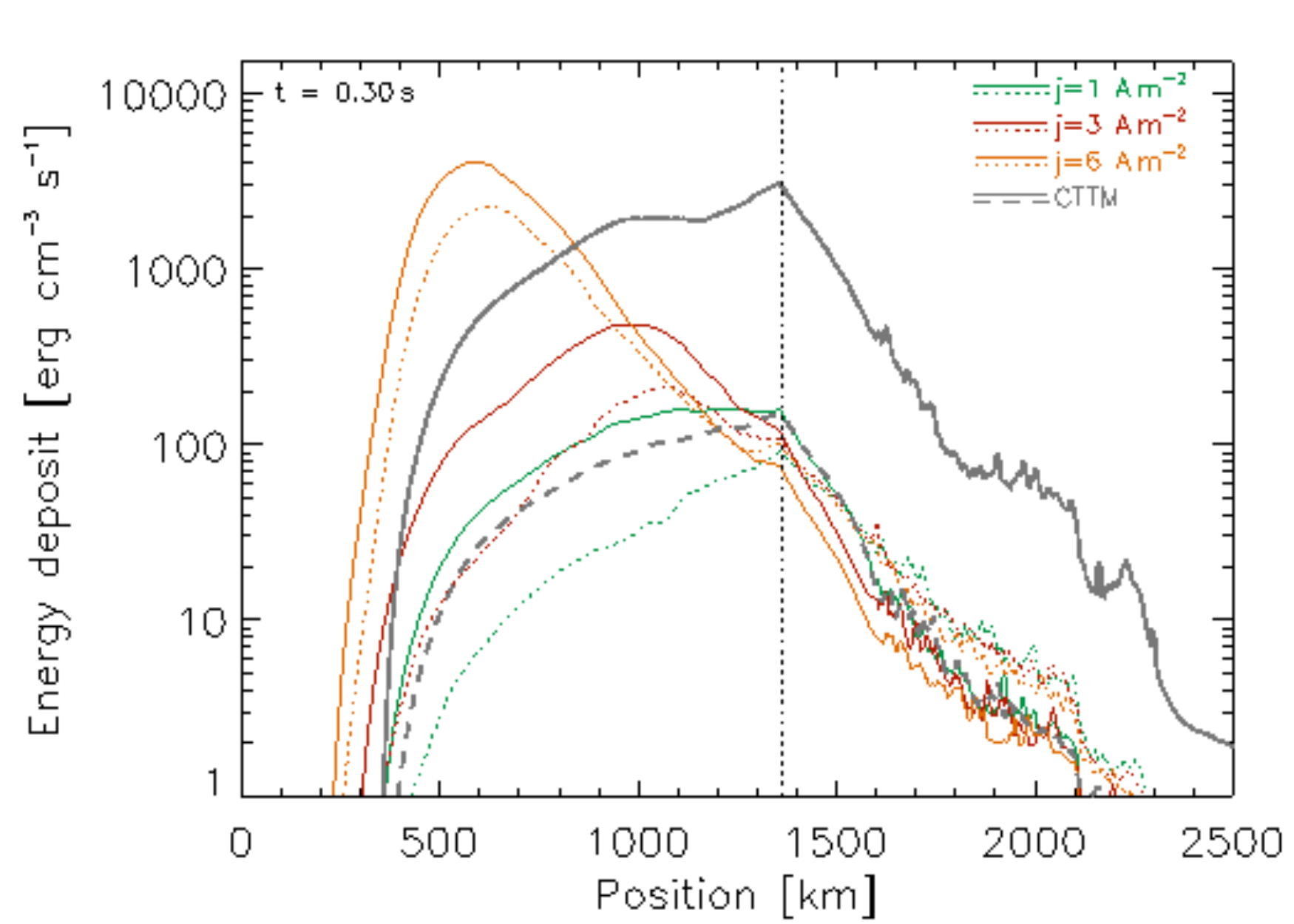}
              \includegraphics[width=8.8cm]{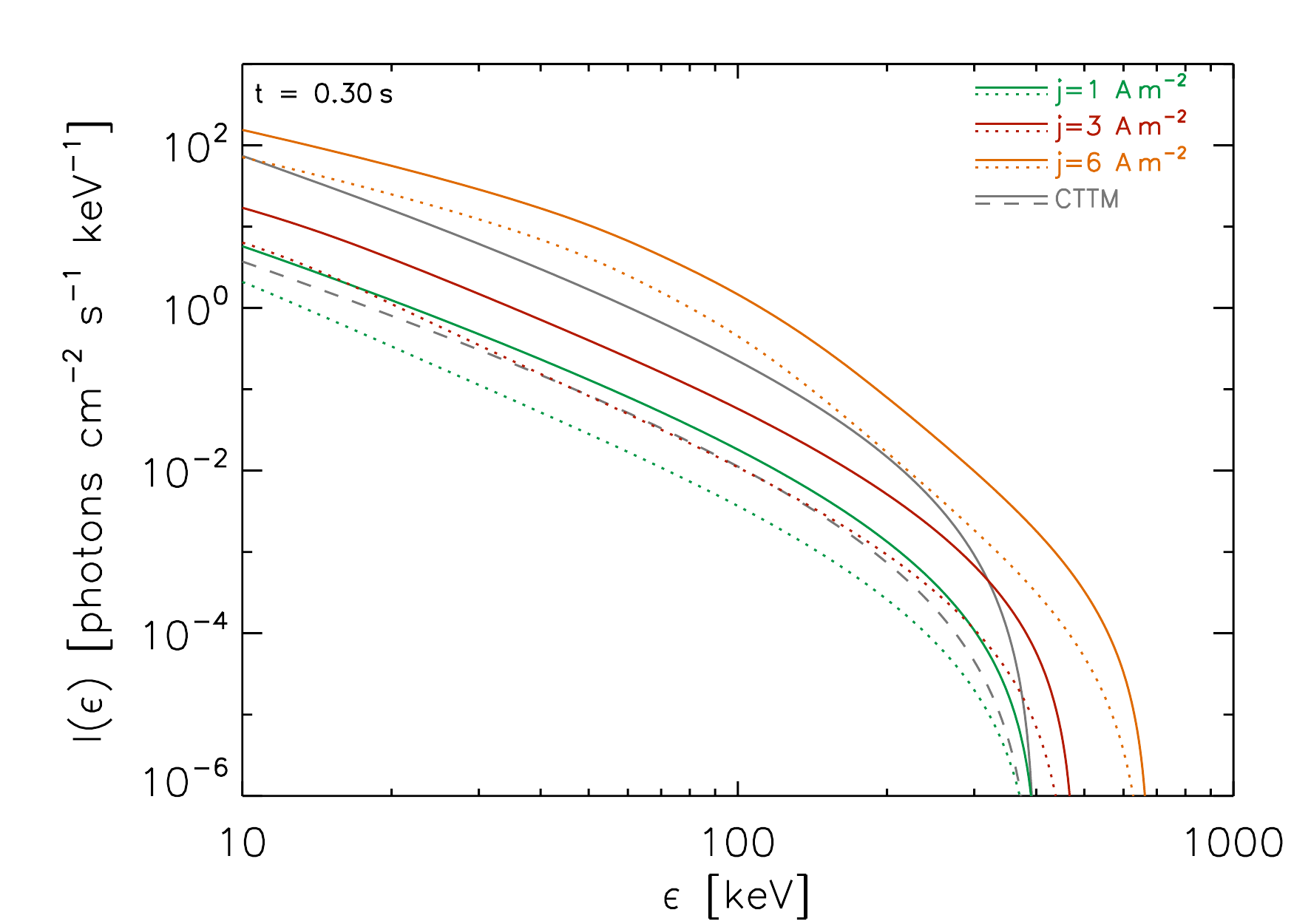}}
 \centerline{\includegraphics[width=8.8cm]{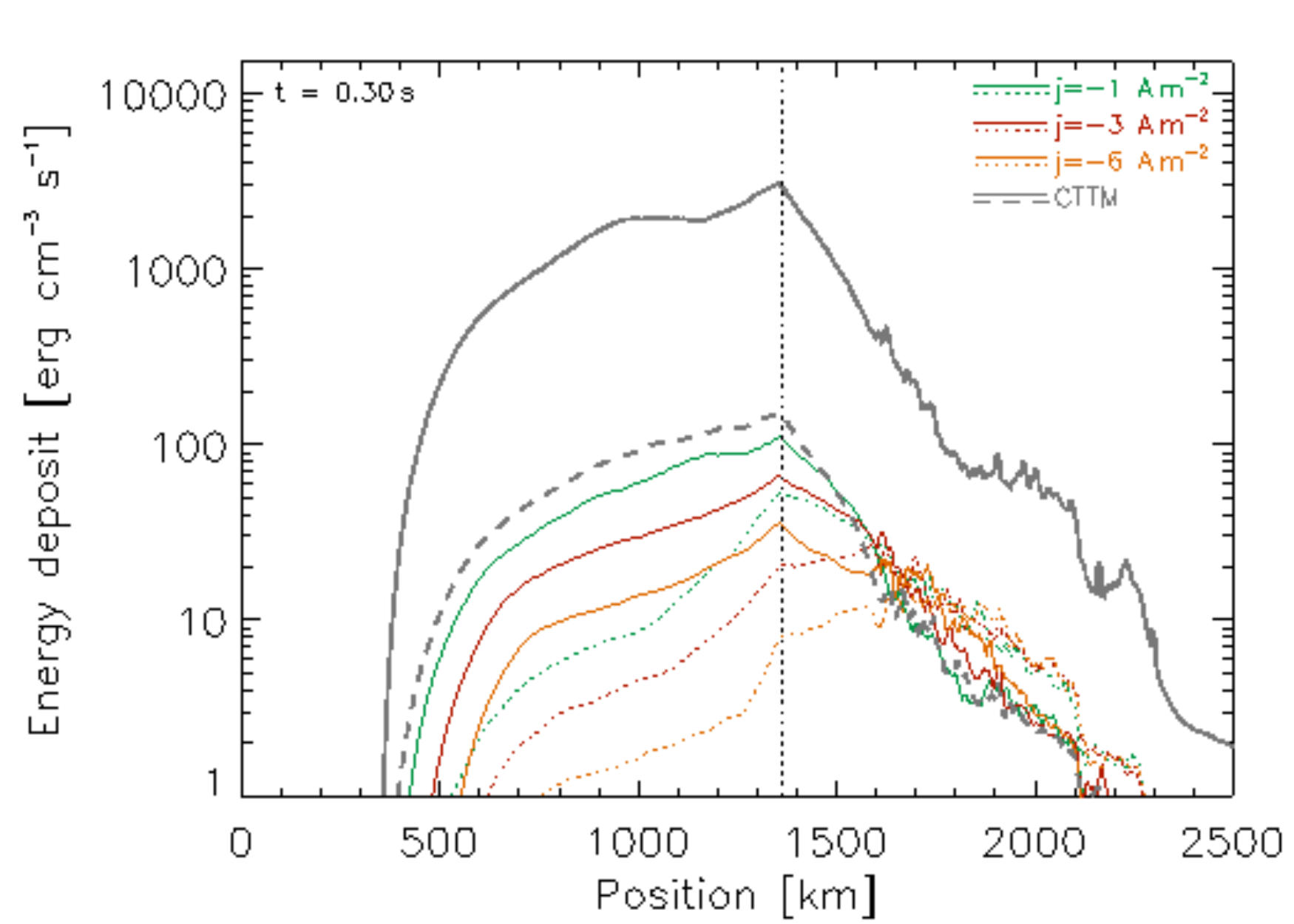}
              \includegraphics[width=8.8cm]{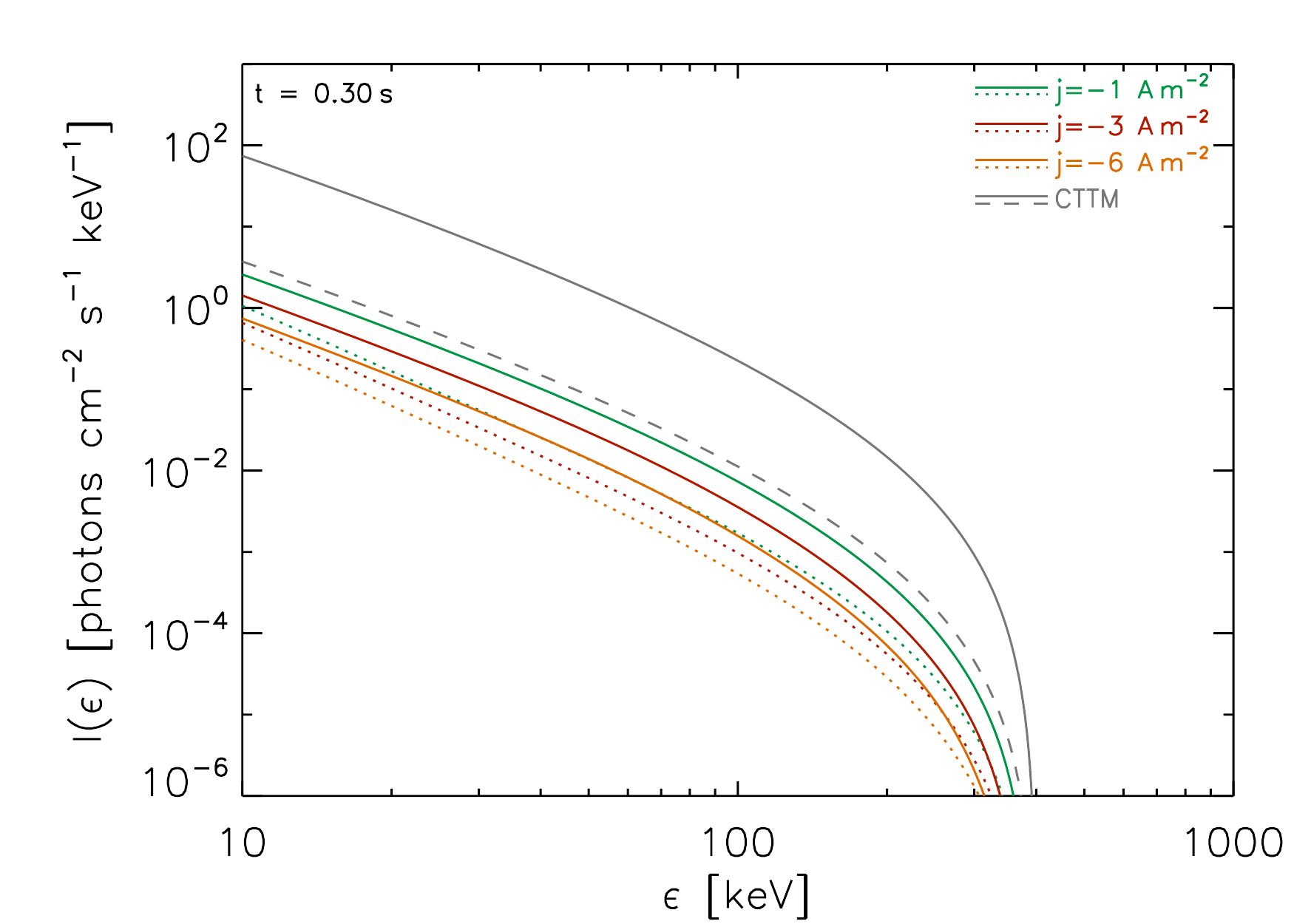}}
\caption{GRTTM instantaneous energy deposits ({\it left})
  and  HXR spectra ({\it right}) for the primary ({\it top}) and
  secondary ({\it bottom}) footpoints and the VAL~C atmosphere at $t=0.3$\,s.
  The green, red, and orange solid ($M^{\mathrm{FF}}$) and dotted
  ($M^{\mathrm{SU}}$) lines correspond to the current densities
  $j=1, 3, 6$\,A\,m$^{-2}$, respectively, and to the energy flux
  $\mathcal{F}_{\mathrm{0}}/2=2.5\times10^{9}$~erg\,cm$^{-2}$\,s$^{-1}$.
  The  grey dashed and solid lines correspond to the CTTM
  with $\mathcal{F}_{\mathrm{0}}/2= 2.5\times10^{9}$ and
  $5\times10^{10}$~erg\,cm$^{-2}$\,s$^{-1}$, respectively.
    The dotted straight vertical line indicates the bottom boundary of the
  magnetic mirror. The HXR spectra are integrated over one half of the loop.}
\label{fig:7}
\end{figure*}

\begin{figure}[!t]
  \centerline{\hspace{0.8em} \includegraphics[width=3.8cm] {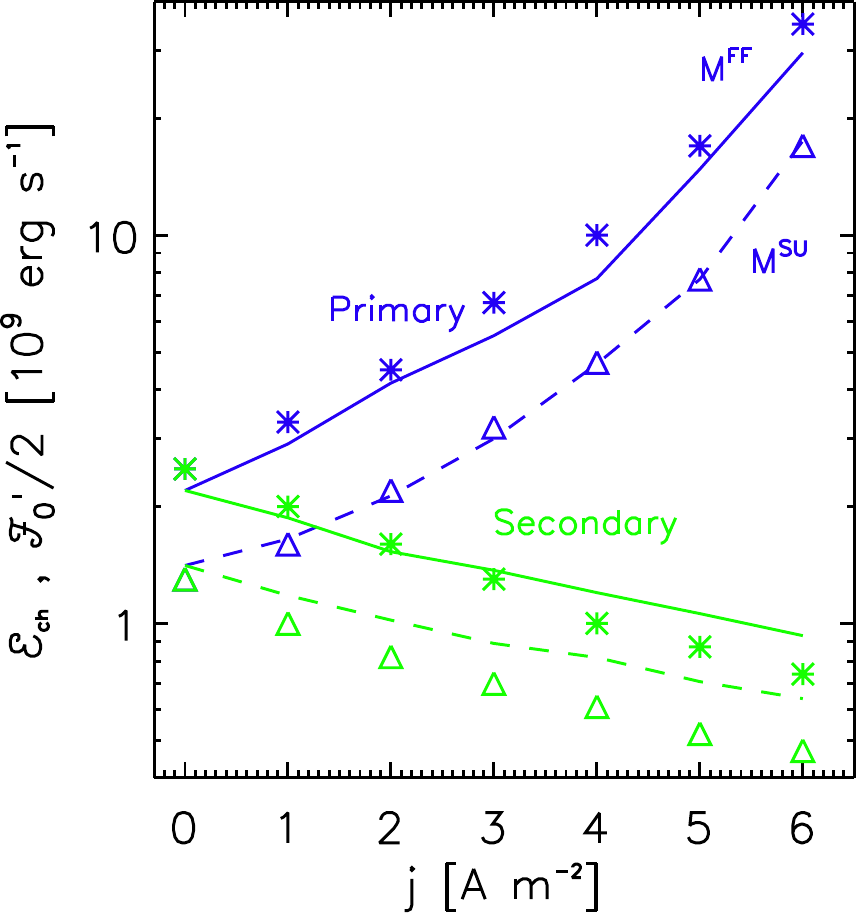}
              \hspace{.3em}  \includegraphics[width=4.45cm]{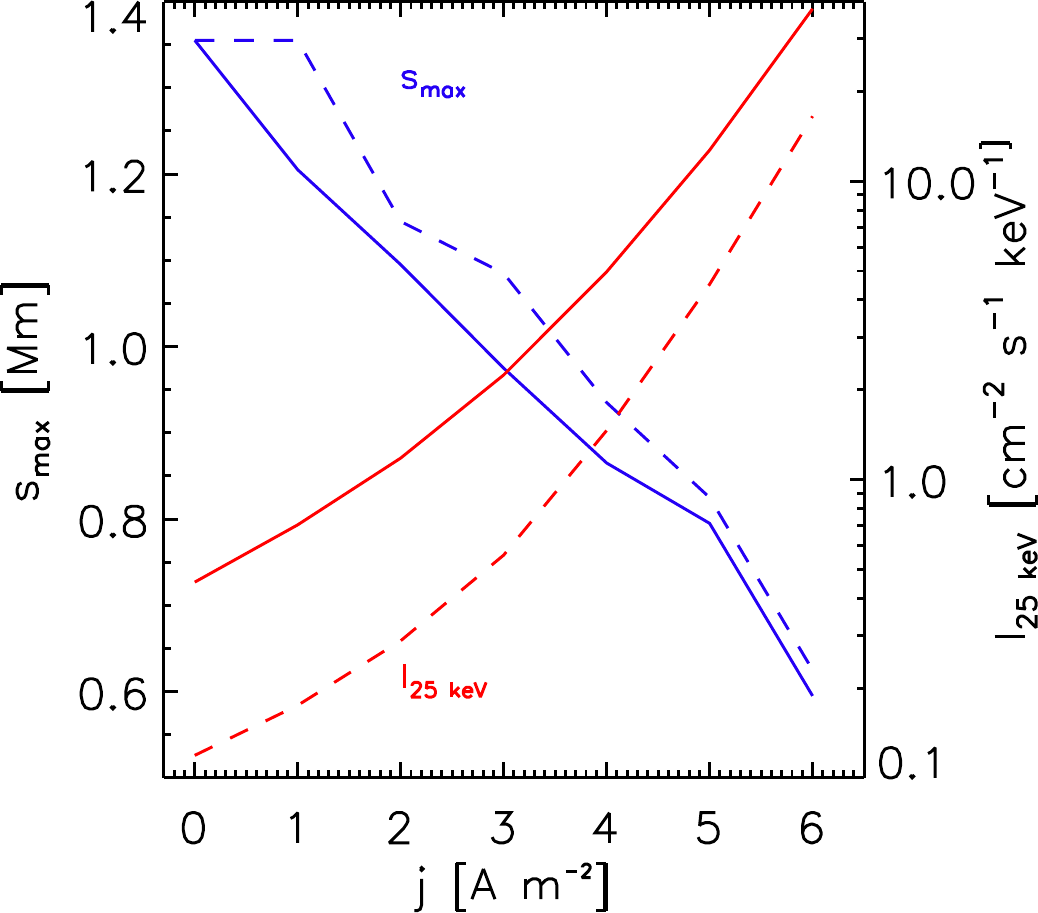}}
  \vspace{0.5em}
  \centerline{               \includegraphics[width=4.05cm]{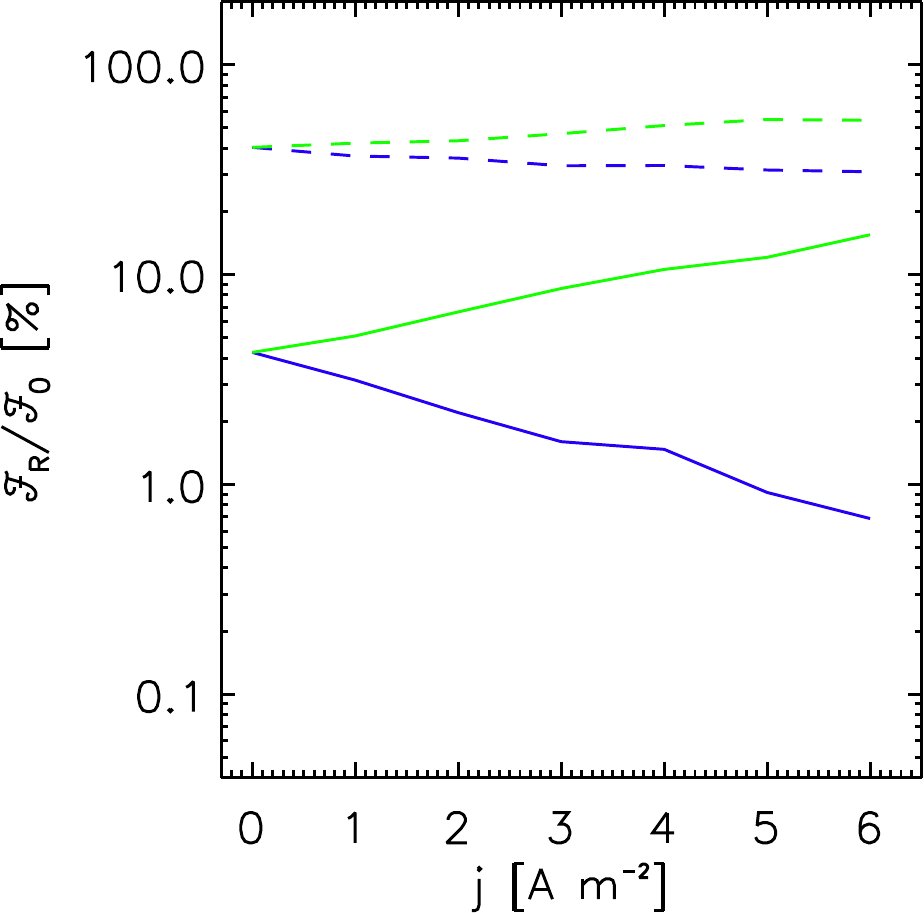}
              \hspace{.8em}  \includegraphics[width=4.25cm]{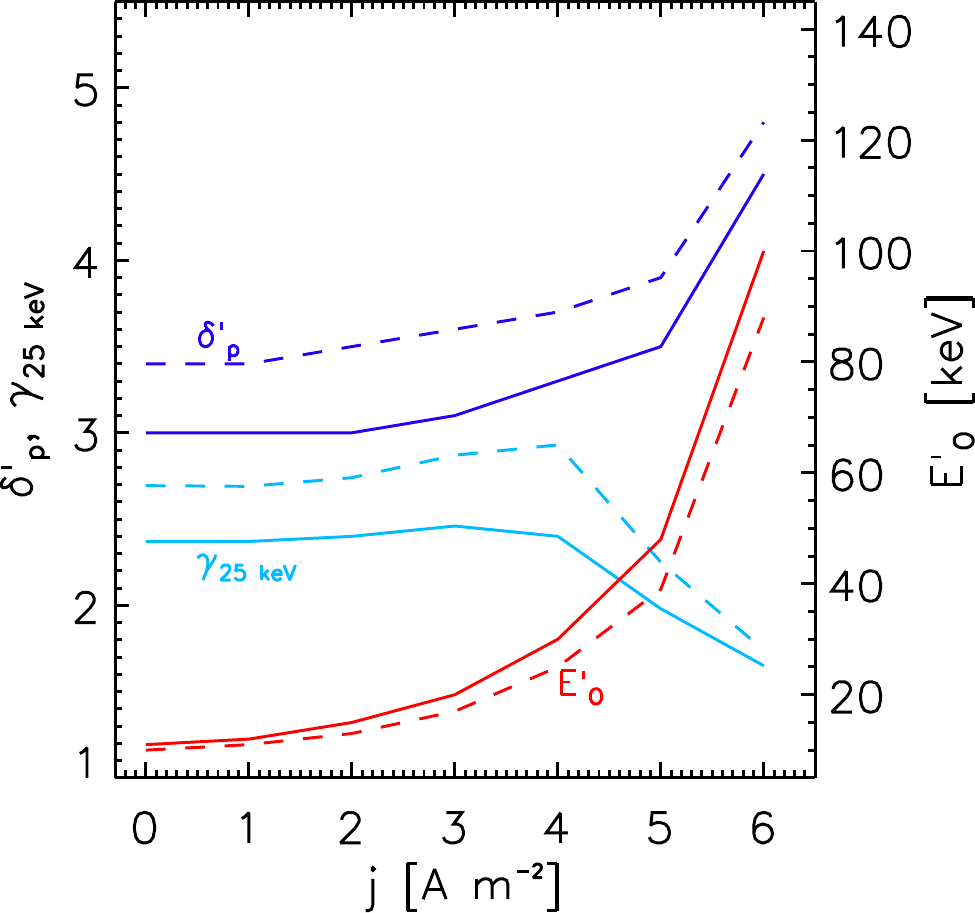}}
\caption{GRTTM summary of calculated parameters of
  chromospheric bombardment for various current
  densities $j$. The solid lines and asterisks denote the
  $M^{\mathrm{FF}}$, the dashed lines and triangles denote the
  $M^{\mathrm{SU}}$. {\it Left:} chromospheric energy deposit
  $\mathcal{E}_{\mathrm{ch}}$ (lines) and fitted energy flux
  $\mathcal{F}_{\mathrm{0}}'$ (symbols) {\it (top)}, the ratio
  $\mathcal{F}_{\mathrm{R}}/\mathcal{F}_{\mathrm{0}}$
  {\it (bottom)}  for the primary (blue) and secondary (green)
  footpoints. {\it Right:} position of energy deposit maximum
  $s_{\max}$ and HXR intensity $I_{25\,\mathrm{keV}}$ ({\it top}),
  HXR spectral index $\gamma_{25\,\mathrm{keV}}$,
  fitted electron beam spectral index $\delta'_\mathrm{p}$ and
  low-energy cutoff $E'_0$ ({\it bottom}) only for the primary
  footpoint.}
  \label{fig:GRTTM_tab}
\end{figure}

The effects of static (global) electric field ${\bf E}_{\mathrm{G}}$
was studied for current densities in the range from 1\,A\,m$^{-2}$ to
6\,A\,m$^{-2}$. The distribution functions of non-thermal electrons
for current  density $j=6$\,A\,m$^{-2}$ and time $t=0.3$\,s after the
beam injection into the loop at its apex are shown in Fig.~\ref{fig:6}.
In the upper left-hand panel, two tails of particles can be identified
in the primary footpoint and the $M^{\mathrm{FF}}$ case.
A faint low-energy tail at energies $E<20$~keV,
located above the bottom boundary of the magnetic mirror, is predominantly
formed of particles with $\mu\le0$ (see the regions labelled L in
Fig.~\ref{fig:6}). Its formation mechanism corresponds to the CTTM, i.e. to
the particle deceleration related to the collisional energy losses in the
target plasma and to the combined effects of particle scattering and magnetic
field convergence, compare with Fig.~\ref{fig:4} (left).
This tail becomes more apparent for distributions that correspond to
lower $j$ (see Fig.~\ref{fig:4}).
On the other hand, a prominent high-energy tail, on energies from 20 to
300~keV stretching from 1.7 to 0.5\,Mm (see the regions labelled
H in Fig.~\ref{fig:6}), does not have any counterpart in Fig.~\ref{fig:4} for
the CTTM. The tail is formed of re-accelerated and relatively focussed particles
with $\mu\approx 1$. Another obvious effects of ${\bf E}_{\mathrm{G}}$ are the
increase in beam penetration depth with growing $j$ and a weakening
 of the population of reflected and back-scattered particles
 propagating towards the secondary footpoint that corresponds
to 0.7\% of the initial beam flux only, see Fig.~\ref{fig:GRTTM_tab}
(left).

Figure~\ref{fig:6} (top right, $M^{\mathrm{SU}}$ case)
exhibits essentially the same features.
The most apparent distinctions between the two distributions
are a much richer population of particles in the
low-energy tail located above the bottom boundary of the magnetic
mirror and the existence of a relatively rich
population of reflected and back-scattered particles with $\mu<0$
(on all energies) propagating towards the secondary
footpoint reaching approximately 30\% of the initial flux
(see Fig.~\ref{fig:GRTTM_tab}, bottom left).
The differences between the distributions corresponding to
$M^{\mathrm{FF}}$ and $M^{\mathrm{SU}}$ cases are solely effects
of the initial $\mu$-distribution.

The situation at the secondary footpoint is shown in
Fig.~\ref{fig:6} (bottom).
In addition to the effect of Coulomb collisions, the field ${\bf
    E}_{\mathrm{G}}$
constantly decreases the parallel velocity component of the particles
propagating towards the secondary footpoint. This results
in the formation of an enhanced low-energy
  tail in the particle distribution functions  located above the
  bottom boundary of the magnetic mirror.
Another obvious feature is a rich population of reflected or
back-scattered particles corresponding approximately
to 15\% and 54\% of the initial beam flux for the $M^{\mathrm{FF}}$
and $M^{\mathrm{SU}}$ cases, respectively (see
Fig.~\ref{fig:GRTTM_tab}, bottom left).
These particles are accelerated by the global field
${\bf E}_{\mathrm{G}}$ back, towards the primary
footpoint.

\onltab{
\begin{table*}[!t]
\caption{Summary of results for the GRTTM with $\mathcal{F}_0/2=2.5\times10^{9}$~erg\,cm$^{-2}$\,s$^{-1}$.}
\label{tab:GRTTM}
\begin{center}
\begin{tabular}{l c c c c c c c c c c c} 
\hline\hline 
Footpoint & ${j}$ & $\mathcal{F_{\mathrm{R}}}/\mathcal{F}_{\mathrm{0}}$  & $\mathcal{E}_{\mathrm{ch}}/10^{9}$ & $s_{\max}$
& $I_{25\,\mathrm{keV}}$ &$\gamma_{25\,\mathrm{keV}}$ & $\mathcal{F}_{\mathrm{0}}'/2\times10^{9}$, $\delta'_\mathrm{p}$, $E'_0$ \\ 

& [A\ m$^{-2}$]& [\%]  & [erg\,s$^{-1}$]  & [Mm] & [cm$^{-2}$\,s$^{-1}$\,keV$^{-1}$] &
& [erg\,cm$^{-2}$\,s$^{-1}$],  , [keV]  \\ 
\hline
          & 1.0 & 3.1 (37)  & 2.8 (1.7) & 1.2 (1.4) & 0.71 (0.18) & 2.4 (2.7) & 3.3, 3.0, 12 (1.6, 3.4, 11) \\
          & 2.0 & 2.2 (36)  & 4.2 (2.1) & 1.1 (1.1) & 1.2 (0.29)  & 2.4 (2.7) & 4.5, 3.0, 15 (2.2, 3.5, 13) \\
Primary& 3.0 & 1.6 (33) & 5.5 (3.0) & 0.98 (1.1) & 2.2 (0.56) & 2.5 (2.9) & 6.7, 3.1, 20 (3.2, 3.6, 17) \\
          & 4.0 & 1.5 (33)  & 7.7 (4.7) & 0.87 (0.94) & 5.0 (1.5) & 2.40 (2.9) & 10, 3.3, 30 (4.7, 3.7, 25) \\
          & 5.0 & 0.92 (32) & 15 (7.7) & 0.80 (0.83) & 13 (4.5)  & 2.0 (2.3) & 17, 3.5, 48 (7.7, 3.9, 39) \\
          & 6.0 & 0.69 (31) & 30 (18) & 0.60 (0.63) & 38 (17) & 1.7 (1.7) & 35, 4.5, 100 (17, 4.8, 88) \\
\hline
          & 1.0 & 5.1 (42) & 1.9 (1.2)  & 1.4 (1.4) & 0.31 (0.086) & 2.4 (2.7) & 2.0, 3.0, 9 (1.0, 3.5, 10) \\
          & 2.0 & 6.6 (43) & 1.5 (1.0)  & 1.4 (1.4) & 0.22 (0.066) & 2.4 (2.7) & 1.6, 3.1, 9  (0.82, 3.5, 10) \\
Secondary & 3.0 & 8.6 (47) & 1.4 (0.89) & 1.4 (1.6) & 0.16 (0.053) & 2.4 (2.7) & 1.3, 3.1, 8 (0.70, 3.5, 10) \\
          & 4.0 & 11 (51) & 1.2 (0.82) & 1.4 (1.6) & 0.13 (0.044) & 2.4 (2.7) & 1.0, 3.1, 8 (0.61, 3.6, 10) \\
          & 5.0 & 12 (55) & 1.1 (0.71) & 1.4 (1.7) & 0.10 (0.034) & 2.4 (2.8) & 0.87, 3.1, 8 (0.52, 3.6, 10) \\
          & 6.0 & 15 (54) & 0.93 (0.64) & 1.4 (1.7) & 0.081 (0.032) & 2.5 (2.8) & 0.74, 3.2, 8 (0.47, 3.6, 10) \\
\hline
\end{tabular}
\end{center}
\end{table*}}

\begin{figure*}
  \centerline{\includegraphics[width=8.8cm]{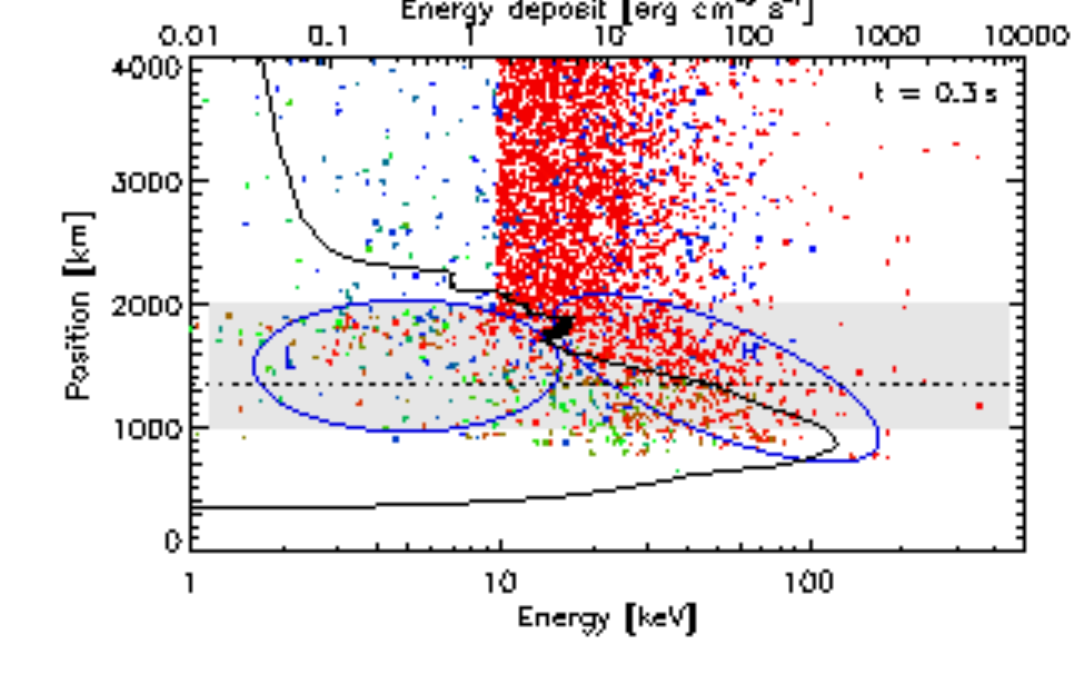}
              \includegraphics[width=8.8cm]{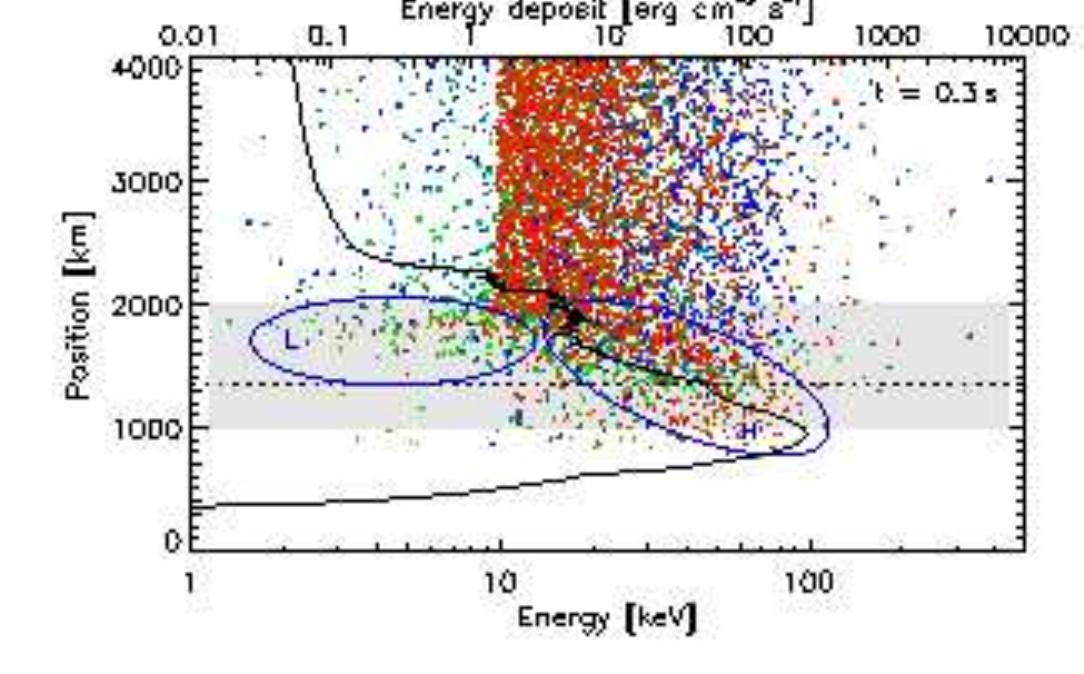}}
\vspace{-2em}
 \centerline{\includegraphics[width=6.cm]{fig_0.pdf}}
\caption{LRTTM ${{\bf E}_{\mathrm{L}}}$-I type distribution
functions of the non-thermal electron energies versus positions
with a colour-coded $M(\mu_0)$ corresponding to
$\overline{E_{\mathrm{L}}} = 0$~V~m$^{-1}$ and
$\mbox{var}(E_{\mathrm{L}}) =1$~V~m$^{-1}$ in the VAL C
atmosphere at time $t=0.3$~s after the  beam injection into the loop at its apex.
{\it Left:} $M^{\mathrm{FF}}$; {\it right:} $M^{\mathrm{SU}}$.
The solid lines indicate the instantaneous energy deposits
corresponding to
$\mathcal{F}_{\mathrm{0}}/2=2.5\times10^{9}$~erg\,cm$^{-2}$\,s$^{-1}$,
the dotted horizontal lines the bottom boundary of the magnetic
mirror, the grey area the secondary re-acceleration region,
and the blue ellipses labelled L and H denote tails in the particle
distribution function. Only the vicinity of the footpoints is
displayed.}
\label{fig:9}
\end{figure*}

The instantaneous energy deposits and HXR spectra for both
the primary and secondary footpoints and various current densities
are shown in Fig.~\ref{fig:7}, and the quantitative results, some
of them only for the primary footpoint, are  summarised in
Fig.~\ref{fig:GRTTM_tab} (see Table~\ref{tab:GRTTM} for complete
results). The magnitudes and spatial distributions of energy deposits
in the atmosphere, as well as the production of HXR photons, are
extremely sensitive to the current densities in the threads.
According to our simulations, the current density $j=6$\,A\,m$^{-2}$
increases $\mathcal{E}_{\mathrm{ch}}$ at the primary footpoint of one
order and $I_{25\,\mathrm{keV}}$ of approximately two orders (see
Fig.~\ref{fig:GRTTM_tab}). Moreover, this HXR spectrum is more
intense than the spectrum of pure CTTM with
$\mathcal{F}_0/2=5\times10^{10}$~erg\,cm$^{-2}$\,s$^{-1}$
(see Fig.~\ref{fig:7}, top right). The
presence of $j$ also considerably changes the distribution of the
energy deposit in the thick-target region. The maximum of the energy
deposit $s_{\max}$ is substantially shifted towards the photosphere
(compare the results for $j=0$ corresponding to
the CTTM and for $j>0$ in the top right of Fig.~\ref{fig:GRTTM_tab}),
and the energy is deposited in a much narrower region in the
chromosphere (see the top left of Fig.~\ref{fig:7}).
In the case of $j=6$\,A\,m$^{-2}$, $\mathcal{E}_{\mathrm{ch}}$ is
comparable to $\mathcal{F}_0/2=5\times10^{10}$~erg\,cm$^{-2}$\,s$^{-1}$
of pure CTTM, however the spatial distribution is completely different.

HXR emission of the primary footpoint comes predominantly
from regions well below the bottom of the
magnetic mirror, close to temperature minimum for
$j \gtrsim3$\,A\,m$^{-2}$ and photon energies $\gtrsim50$~keV.
As $j$ increases, HXR spectra  get more intense and flatter at
deka-keV energies, and the maximum photon energy is shifted to
higher energies. This is all consistent with the presence of the
high-energy electrons accelerated by ${\bf E}_{\mathrm{G}}$ below
the magnetic mirror. Although the HXR power-law index
$\gamma_{25\,\mathrm{keV}}$ tends to harden as $j$ increases,
the fitted CTTM injected electron power-law index
$\delta_{\mathrm{p}}'$ becomes steeper. However, at the same time,
the low-energy cutoff $E_0'$ rises to deka-keV values, causing
decrease in $\gamma_{25\,\mathrm{keV}}$ --~see fitted parameters
in Fig.~\ref{fig:GRTTM_tab} (bottom right).

The model of $j=1$\,A\,m$^{-2}$ is similar to the CTTM situation; i.e. similar formation heights of HXR,
spectral shape of photon spectrum (Fig.~\ref{fig:7}, left),
and fitted electron distribution  (Fig.~\ref{fig:GRTTM_tab}, bottom right).
In the case of $j=6$\,A\,m$^{-2}$, the HXR spectra are
extremely flat below $\sim$40~keV with $E_0'\sim 100$~keV.
Such low-energy cutoffs are not found from observations, therefore this case could represent a
limit of possible $j$ in flare loops.

\begin{figure*}[!t]
  \centerline{\includegraphics[width=8.cm]{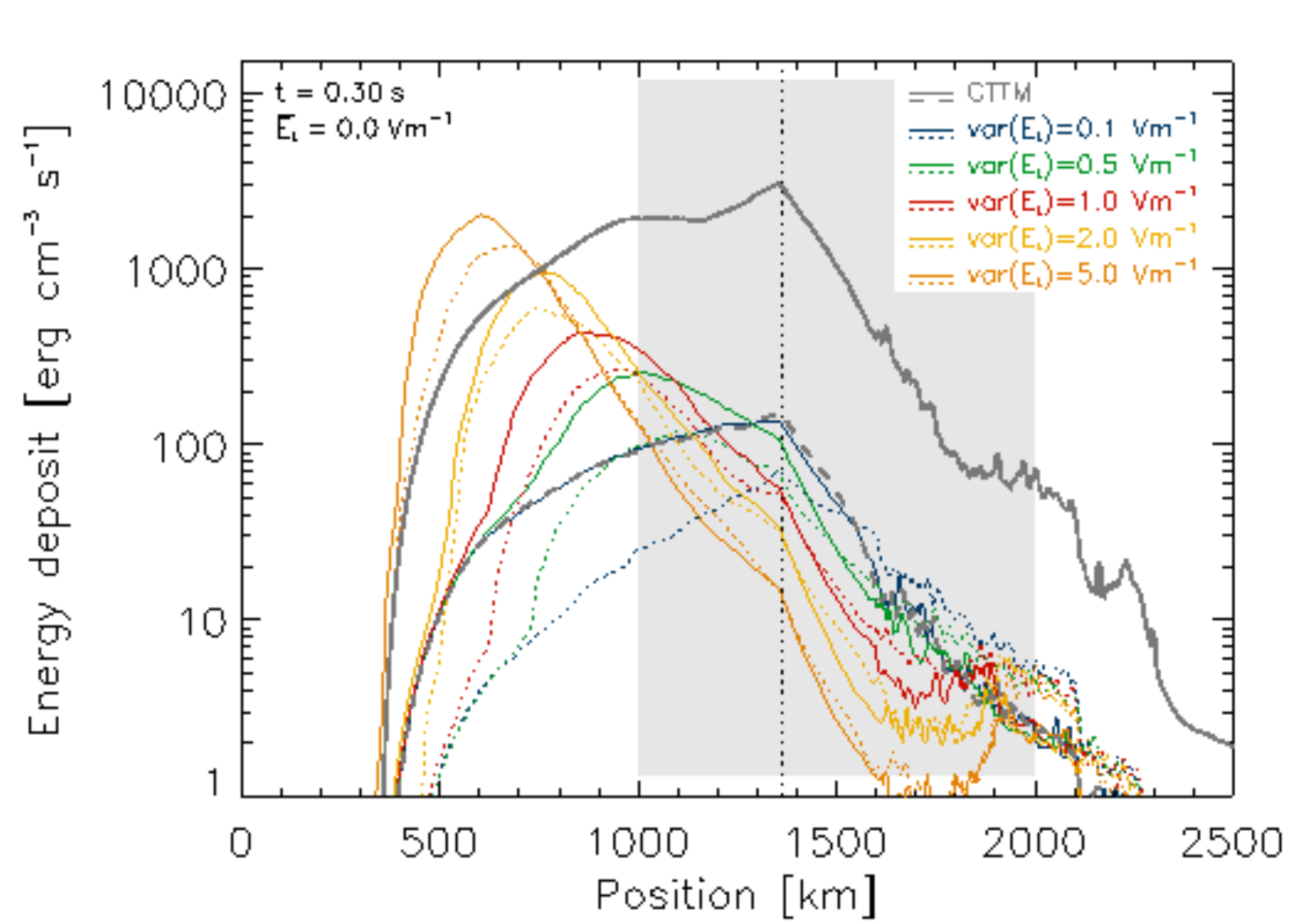}
              \includegraphics[width=8.cm]{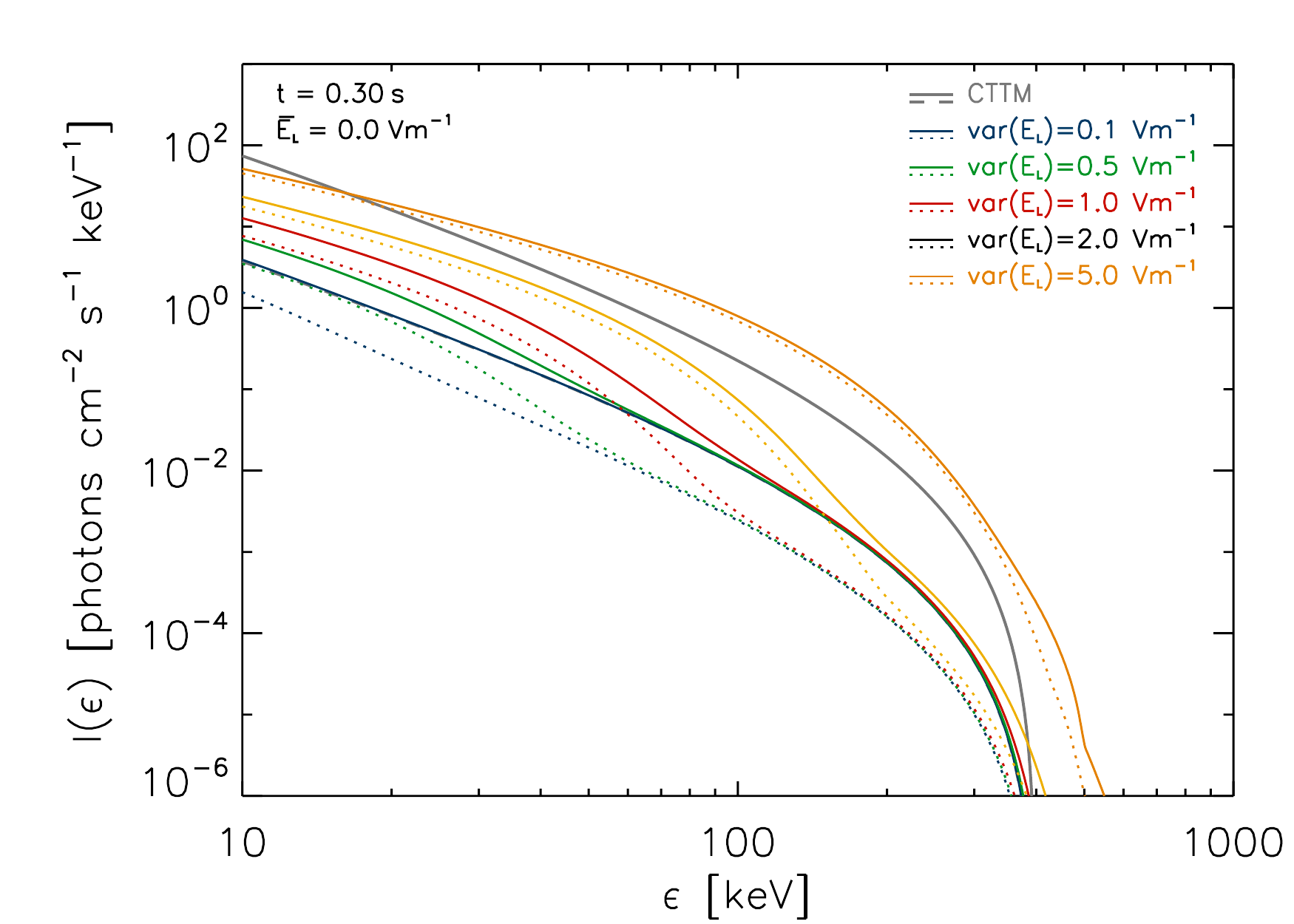}}
\caption{LRTTM ${{\bf E}_{\mathrm{L}}}$-I type
  instantaneous energy deposits ({\it left}) and HXR spectra
  ({\it right}) for the VAL~C  atmosphere at $t=0.3$~s. The solid ($M^{\mathrm{FF}}$) and
  dotted ($M^{\mathrm{SU}}$) blue,
  green, red, yellow, and orange lines correspond to $\overline{E_{\mathrm{L}}}=0$~V\,m$^{-1}$ and
  $\mbox{var}(E_{\mathrm{L}})=0.1, 0.5, 1, 2, 5$~V\,m$^{-1}$, respectively,
  and energy flux $\mathcal{F}_{\mathrm{0}}/2=2.5\times10^{9}$~erg\,cm$^{-2}$\,s$^{-1}$.
  The dashed and solid grey lines correspond to the CTTM
  with $\mathcal{F}_{\mathrm{0}}/2= 2.5\times10^{9}$ and
  $5\times10^{10}$~erg\,cm$^{-2}$\,s$^{-1}$, respectively.
 The dotted straight vertical line indicates the
  bottom boundary of the magnetic mirror, the grey area
  the secondary re-acceleration region. The HXR
  spectra are integrated over one half of the loop.}
 \label{fig:10}
\end{figure*}

The situation at the secondary footpoint is different (see
Fig.~\ref{fig:7}, bottom). Because a part of energy carried
by non-thermal particles is drained due to the actuation of
${\bf E}_{\mathrm{G}}$, the resulting chromospheric energy
deposits for a particular $j$ are smaller than at the primary
footpoint. As expected, this behaviour steeply increases with $j$.
Although the HXR spectra of the secondary footpoint are less intense than the
spectrum of pure CTTM, the overall spectral shape is not changed significantly.
Consequently, the fitted injected electron beam parameters show only a decrease
of $\mathcal{F}_{\mathrm{0}}'$ consistent with lower
$\mathcal{E}_{\mathrm{ch}}$ (see Fig.~\ref{fig:GRTTM_tab}, top left)
and  Table~\ref{tab:GRTTM}.

\onltab{
\begin{table*}[!t]
\caption{Summary of results for the LRTTM with $\mathcal{F}_0/2=2.5\times10^{9}$~erg\,cm$^{-2}$\,s$^{-1}$.}
\label{tab:LRTTM}
\begin{center}
\begin{tabular}{c c l l c c c l}
\hline\hline
 $\overline{E_{\mathrm{L}}}$ & $\mbox{var}({E_{\mathrm{L}}})$ &
 $\mathcal{F_{\mathrm{R}}}/\mathcal{F}_{\mathrm{0}}$ &$\mathcal{E}_{\mathrm{ch}}/10^{9}$ &
 $s_{\max}$ & $I_{25\,\mathrm{keV}}$ &$\gamma_{25\,\mathrm{keV}}$ &
 $\mathcal{F}_{\mathrm{0}}'/2\times10^{9}$, $\delta'_\mathrm{p}$, $E'_0$ \\

[V\,m$^{-1}$] & [V\,m$^{-1}$] & [\%] & [erg\,s$^{-1}$] & [Mm] & [cm$^{-2}$\,s$^{-1}$\,keV$^{-1}$] &
& [erg\,cm$^{-2}$\,s$^{-1}$, , keV] \\
\hline
0.0   & 0.1 & 5.8 (41) & 2.2 (1.4) & 1.3  (1.4) & 0.46 (0.12) & 2.4 (2.7) & 2.2, 3.0, 12 (1.2, 3.5, 11) \\
      & 0.5 & 21 (78) & 3.0 (1.7) & 1.0  (1.1) & 0.79 (0.32) & 2.8 (3.3) & 2.9, 3.7, 24 (1.5, 4.4, 25) \\
      & 1.0 & 31 (120)  & 3.9 (2.5) & 0.87 (0.96)  & 2.0 (1.2)   & 2.4 (2.5) & 4.1, 4.4, 41 (2.6, 4.9, 42) \\
      & 2.0 & 44 (130)  & 6.4 (4.5) & 0.78 (0.76)  & 4.8 (3.6)   & 1.9 (1.9) & 6.2, 4.8, 70 (4.7, 5.0, 71) \\
      & 3.0 & 68 (140)  & 7.0 (5.5) & 0.69 (0.71)  & 7.5 (6.0)   & 1.8 (1.8) & 7.9, 4.6, 93  (6.4, 4.8, 93) \\
      & 4.0 & 77 (170)  & 9.1 (7.5) & 0.66 (0.70)  & 10 (8.3)   & 1.7 (1.7) & 9.7, 4.5, 110  (7.9, 4.6, 110) \\
      & 5.0 & 90 (160)  & 12 (8.6) & 0.61 (0.68)  & 13 (11)   & 1.6 (1.6) & 11, 4.3, 130 (9.8, 4.4, 130) \\
\hline
0.1   & 0.0 & 0.05 (23) & 7.5 (6.5) & 0.91 (0.94) & 4.2 (2.8) & 2.4 (2.6) & 8.7, 6.0, 47 (6.5, 9.0, 48) \\
      & 0.5 & 0.63 (37) & 8.3 (6.1) & 0.88 (0.89) & 4.7 (3.3) & 2.3 (2.40) & 8.7, 5.7, 51 (6.6, 7.0, 52) \\
      & 1.0 & 3.2 (55) & 8.7 (8.2) & 0.82 (0.84) & 6.2 (4.6) & 2.1 (2.1) & 9.5, 5.6, 61 (7.3, 6.0, 61) \\
      & 2.0 & 17 (78) & 10 (8.7) & 0.72 (0.74) & 8.7 (7.3) & 1.9 (1.9) & 10, 5.8, 84  (8.6, 5.8, 85) \\
      & 3.0 & 37 (100)  & 12 (8.7) & 0.68 (0.70) & 11 (9.8) & 1.7 (1.7) & 11, 4.8, 100  (9.9, 5.0, 110) \\
      & 4.0 & 65 (120)  & 9.0 (12) & 0.63 (0.63) & 13 (12) & 1.6 (1.6) & 12, 4.6, 120  (11, 4.7, 130) \\
      & 5.0 & 76 (150)  & 13 (11) & 0.63 (0.65) & 16 (14) & 1.6 (1.6) & 13, 4.4, 140 (12, 4.4, 140) \\
\hline 
\end{tabular}
\end{center}
\end{table*}}

\subsection{LRTTM}

\subsubsection{${{\bf E}_{\mathrm{L}}}$-I type}

The non-thermal electron distribution functions for the stochastic field with
$\overline{E_{\mathrm{L}}}=0$~V\,m$^{-1}$ and
$\mbox{var}({E}_{\mathrm{L}})=1$~V\,m$^{-1}$ in the VAL~C atmosphere and  time
$t=0.3$~s after the beam injection into the loop at its apex are shown in
Fig.~\ref{fig:9}. In both panels two kinds of particle populations can be
identified: a conspicuous high-energy tail fuzzy in energies at
particular height sections (see the regions labelled H), and an inconspicuous
low-energy tail (see the regions labelled L).

The high-energy tail is located within the
re-acceleration region on energies from 10 to 100~keV. It
indicates that the net re-acceleration of particles occurs even though
any electron in the re-acceleration region has an equal
probability of encountering stochastic field ${{\bf E}_{\mathrm{L}}}$ (normally distributed)
of parallel or anti-parallel orientation relative to $\mu=1$. The net
acceleration in this type of electric field is a consequence
of inverse proportionality between the electron collisional energy
loss and energy $\mbox{d}E/\mbox{d}z\propto1/E$,
$z$ being the column density  \citep{emslie1978}. The energy gain of re-accelerated
electrons increases with $\mbox{var}({E}_{\mathrm{L}})$ similar to
the fuzziness of the high-energy tails and the fluxes of backwards
moving electrons (with $\mu<0$).
The ratio $\mathcal{F_{\mathrm{R}}}/\mathcal{F}_{\mathrm{0}}$ corresponding to
$\mbox{var}({{E}_{\mathrm{L}}})=1$~V\,m$^{-1}$ is approximately 31\%
and 120\% for the $M^{\mathrm{FF}}$  and $M^{\mathrm{SU}}$ cases,
respectively; i.e., in the latter case the backward energy flux exceeds
the initial flux propagating downwards from the corona
(see Fig.~\ref{fig:LRTTM_a_new}, right). Another effect of growing
$\mbox{var}{({E}_{\mathrm{L}})}$ is a decrease in the electron population
 having $\mu$ distinct from 1 or $-1$.
Ultimately, for high values of
$\mbox{var}{({E}_{\mathrm{L}})}$, only particles with $\mu$ either close to 1 or $-1$
are present in the distribution, so the $\mu$-distribution then copies the
directional distribution of the re-accelerating field.

The inconspicuous low-energy tail spreads from the top of
the re-acceleration region to the lower boundary of the magnetic
mirror, and it is formed of particles of all possible pitch angles with
energies under 20~keV. It is shifted higher into the chromosphere in
comparison to the low-energy tail
in the CTTM case (see Fig.~\ref{fig:4}). As $\mbox{var}{({E}_{\mathrm{L}})}$
increases, the low-energy tail becomes less distinct and its location
is shifted higher towards the upper boundary of the re-acceleration region.
The low-energy tail is formed
by concerted actuation of Coulomb collisions and alternating stochastic field.

The energy deposits and HXR spectra corresponding to various values of
$\mbox{var}(E_{\mathrm{L}})$ in the range from 0.1 to 5~V\,m$^{-1}$ are
shown in Fig.~\ref{fig:10} and their main parameters $\mathcal{E}_{\mathrm{ch}}$, $s_{\max}$,
$I_{25\,\mathrm{keV}}$, and $\gamma_{25\,\mathrm{keV}}$ are
displayed in
left-hand panels of Figs.~\ref{fig:LRTTM_a_new} and \ref{fig:LRTTM_b_new}
and summarised in Table~\ref{tab:LRTTM}.

The behaviour of the energy deposits is similar to the GRTTM of the
primary footpoint. They increase with $\mbox{var}{({E}_{\mathrm{L}})}$,
$s_{\max}$ are shifted to the deeper layers, and the energy is
deposited into an even narrower chromospheric region. For the lowest
studied value $\mbox{var}(E_{\mathrm{L}})=0.1$~V\,m$^{-1}$, we obtained
practically no change in all followed parameters relative to the CTTM with an
identical initial flux (see Figs.~\ref{fig:10}, \ref{fig:LRTTM_a_new} 
and \ref{fig:LRTTM_b_new}).

On the other hand, for the maximum value
$\mbox{var}(E_{\mathrm{L}}) =5$~V\,m$^{-1}$ there is half an
order increase in $\mathcal{E}_{\mathrm{ch}}$ and
a substantial shift of $s_{\max}$ towards the photosphere ($\sim$750~km)
for both initial $\mu$-distributions. The value of $I_{25\,\mathrm{keV}}$
increases considerably  ($28\times$ for the $M^{\mathrm{FF}}$
and $10^2\times$ for the $M^{\mathrm{SU}}$ case) relative to the
CTTM  with an identical initial flux (see Fig.~\ref{fig:LRTTM_b_new},
left).

Again, hard X-ray emission comes from the regions below the magnetic mirror.
As for GRTTM case, as $\mbox{var}(E_{\mathrm{L}})$ increases, the LRTTM
hard X-ray spectra at $\sim$25~keV become flatter (see
$\gamma_{25\,\mathrm{keV}}$ in Fig.~\ref{fig:LRTTM_b_new}, left).
Values of $\mbox{var}(E_{\mathrm{L}}) \ge
2$~V\,m$^{-1}$ result in extremely flat photon spectra. On the other hand, the
LRTTM X-ray spectra exhibit a double break or a local sudden decrease; see e.g.
the spectrum in the $\sim$50~--~100~keV range corresponding to
$\mbox{var}(E_{\mathrm{L}})=1.0$~V\,m$^{-1}$ in Fig.~\ref{fig:10}, right. Such
spectral shapes affect the fitted CTTM electron distributions and result in
high values of $E'_0$ (located approximately at the energy of a double break)
and higher values of $\delta'_0$ (see Fig.~\ref{fig:LRTTM_b_new}, bottom left).
As $\mbox{var}(E_{\mathrm{L}})$ rises, $E'_0$ still increases but $\delta_0'$
stays almost constant, i.e.~4~--~5. The model of
$\mbox{var}(E_{\mathrm{L}})=5$~V\,m$^{-1}$ presents a limit, and the hard X-ray
spectrum is consistent with a rather flat electron flux spectrum of high
$E'_0$. Although the spectrum is more intense than the spectrum of pure CTTM
with $\mathcal{F}_{\mathrm{0}}/2=5\times10^{10}$~erg\,cm$^{-2}$\,s$^{-1}$
(i.e. 20$\times$ higher than the initial flux used in this model), owing to the
high value of $E'_0$, the fitted electron flux is lower and consistent with the
energy deposit in the chromosphere $\mathcal{E}_\mathrm{ch}$ (see
Fig.~\ref{fig:LRTTM_a_new}, left).

\subsubsection{${{\bf E}_{\mathrm{L}}}$-II type}

\begin{figure}[!t]
  \centerline{ \includegraphics[width=4.1cm] {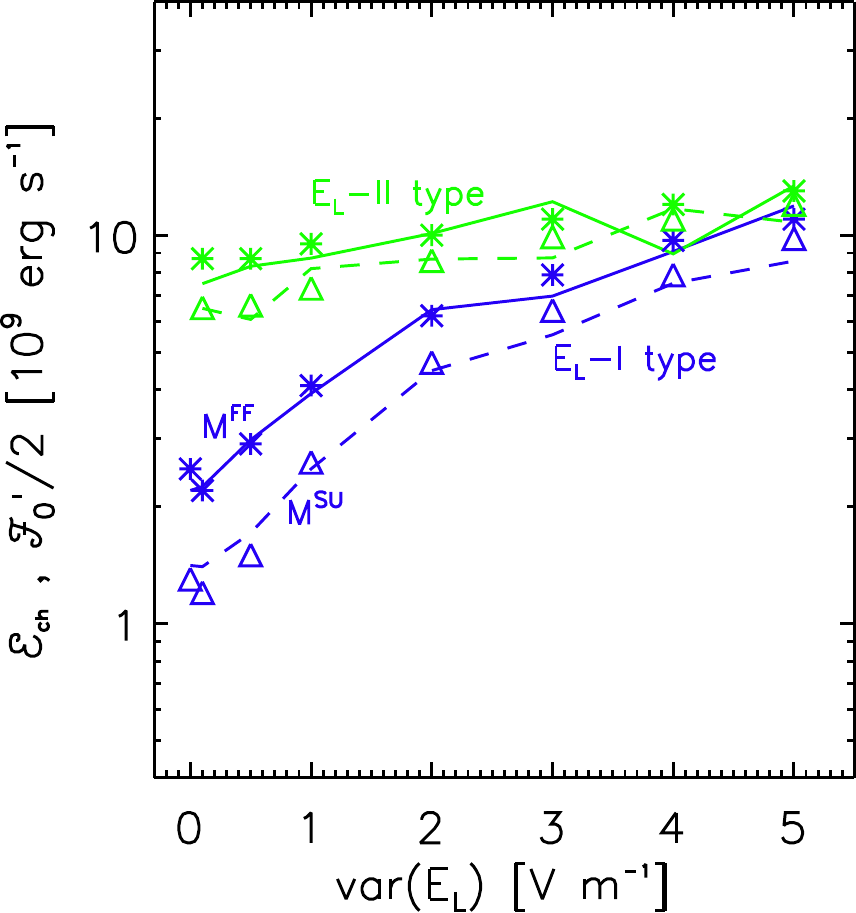}
  \hspace{.3em}\includegraphics[width=4.35cm]{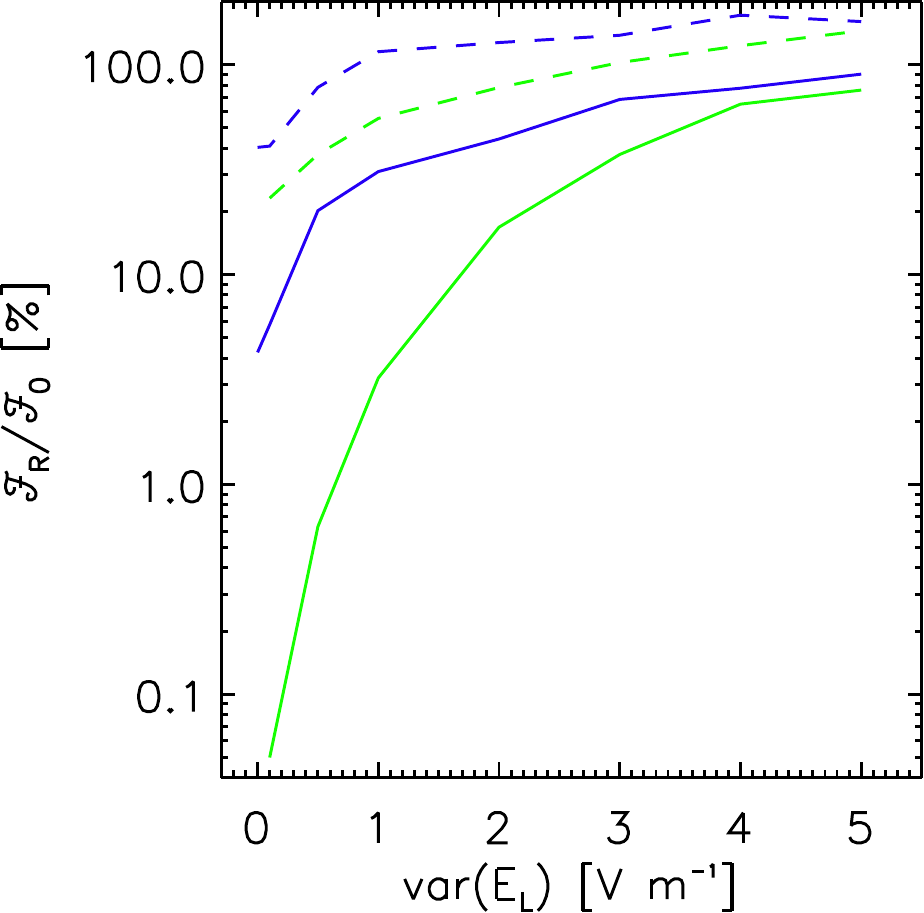}}
 \caption{LRTTM ${\bf E}_{\mathrm{L}}$-I (blue) and  ${\bf
     E}_{\mathrm{L}}$-II (green) (for
        $\overline{E_{\mathrm{L}}}=0.1$\,V\,m$^{-1}$)  chromospheric energy deposits
     $\mathcal{E}_{\mathrm{ch}}$ (lines) and
     fitted energy flux $\mathcal{F}_{\mathrm{0}}'$ (symbols) ({\it
     left}) and the ratio $\mathcal{F}_{\mathrm{R}}/\mathcal{F}_{\mathrm{0}}$
    ({\it right}) for various $\mbox{var}(E_{\mathrm{L}})$. Solid
    lines and asterisks denote $M^{\mathrm{FF}}$;
    dashed lines and triangles denote $M^{\mathrm{SU}}$.}
 \label{fig:LRTTM_a_new}
\end{figure}

The effects of local re-acceleration due to the stochastic field
${{\bf E}_{\mathrm{L}}}$ with $\overline{E_{\mathrm{L}}}\neq 0$ are
demonstrated for the case with $\overline{E_{\mathrm{L}}}=0.1$~V\,m$^{-1}$ and
$\mbox{var}(E_{\mathrm{L}})=0.5$~V\,m$^{-1}$ (see the distribution
functions for
$M^{\mathrm{FF}}$ and $M^{\mathrm{SU}}$ cases in Fig.~\ref{fig:13}).
The re-acceleration process again results
in formation of fuzzy high-energy tail of particles situated in the
secondary acceleration region and covering the energy range from 10 to
100~keV approximately (see the regions labelled H). The mean
energy reached by the re-accelerated
electrons at the lower boundary of the re-acceleration region steeply
increases with $\overline{E_{\mathrm{L}}}$, and at the same time the
maximum of energy deposit shifts towards the deeper layers. The mean
value of ${{\bf E}_{\mathrm{L}}}$ also
has a strong focussing effect on the re-accelerated electrons.
The latter effect reduces the ratio of backscattered and reflected
particle flux to the initial flux $\mathcal{F_{\mathrm{R}}}/\mathcal{F}_{\mathrm{0}}$
to less than 1\% for the $M^{\mathrm{FF}}$ and to 37\% for the $M^{\mathrm{SU}}$
case, respectively: compare values
of $\mathcal{F_{\mathrm{R}}}/\mathcal{F}_{\mathrm{0}}$  for the
individual field types and parameters of ${{\bf E}_{\mathrm{L}}}$
displayed in Fig.~\ref{fig:LRTTM_a_new} (right). The value of
$\mbox{var}(E_{\mathrm{L}})$ plays a similar role to what is  described
above for the ${{\bf E}_{\mathrm{L}}}$-I type. In comparison with the
effects of $\overline{E_{\mathrm{L}}}$, it only weakly influences the
energy gain of electrons
at the lower boundary of the re-acceleration region, it increases the
fuzziness of the high-energy tail and the flux of backwards moving
electrons (with $\mu<0$). For high values of
$\mbox{var}(E_{\mathrm{L}})$ we also see a decrease in electrons
having $\mu$ other than close to 1 and $-1$, which is again the effect of
imprint of the directional distribution of  ${{\bf E}_{\mathrm{L}}}$
on the electron $\mu$-distribution, which was also found for the
stochastic field type ${{\bf E}_{\mathrm{L}}}$-I.

\begin{figure}[!t]
    \centerline{ \includegraphics[width=4.cm]  {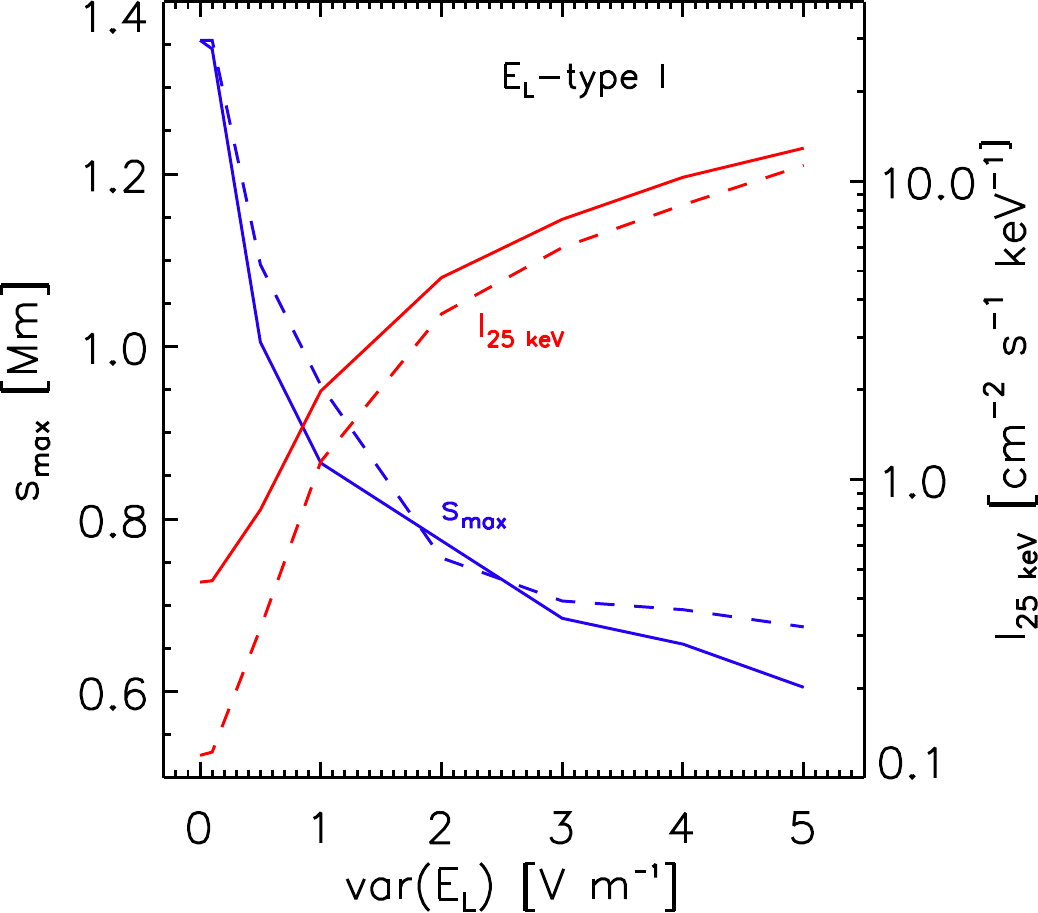}
    \hspace{.3em}\includegraphics[width=4.cm]  {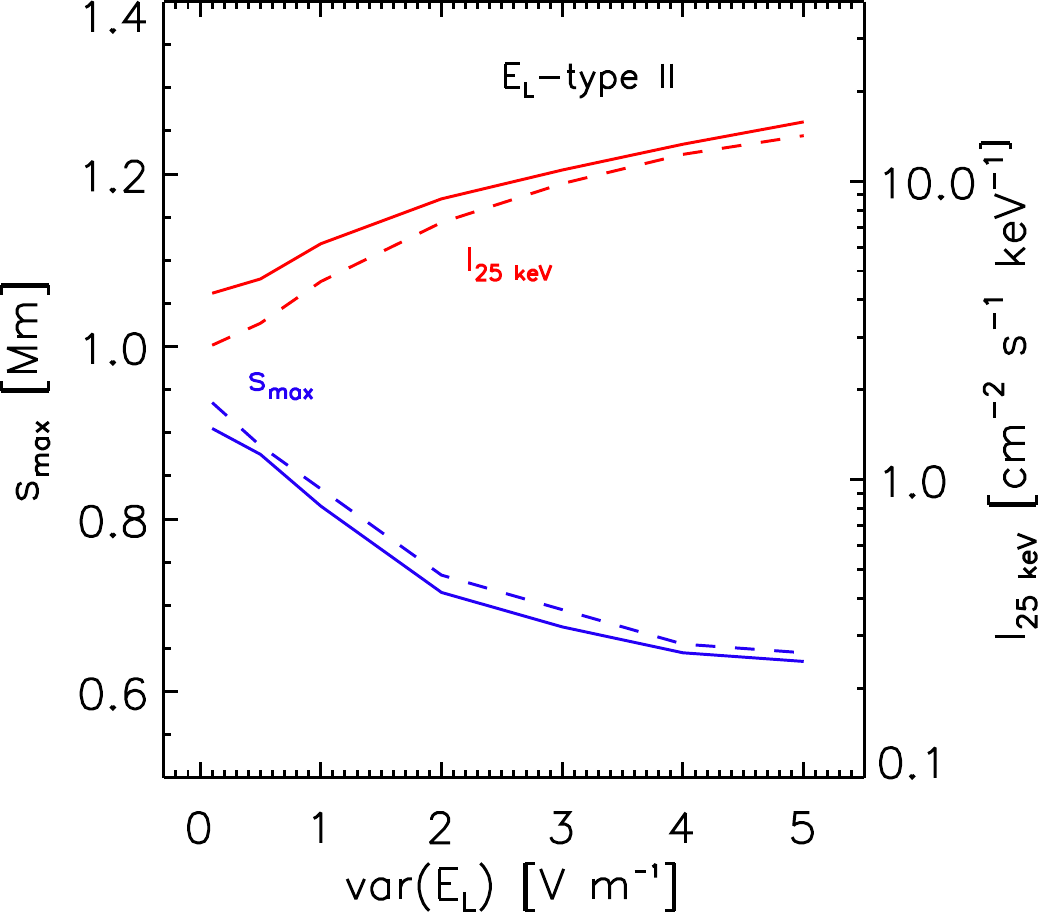}}
    \vspace{0.5em}
    \centerline{ \includegraphics[width=4.25cm]{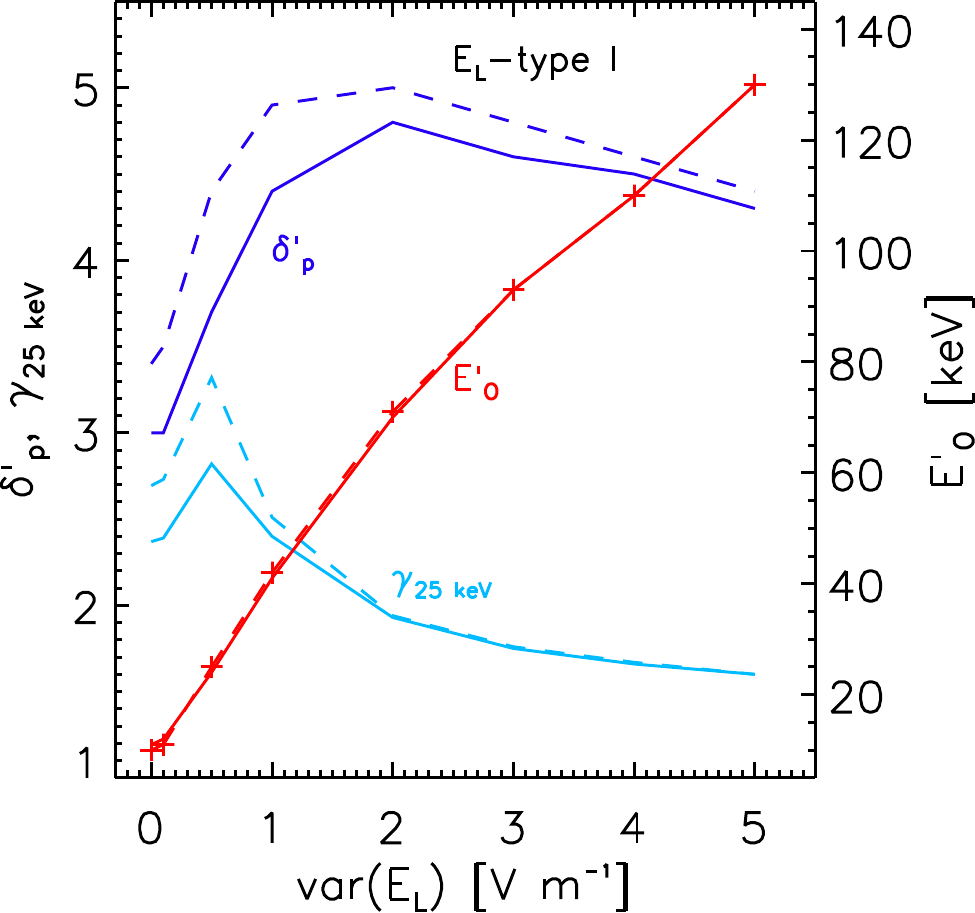}
    \hspace{.3em}\includegraphics[width=4.25cm]{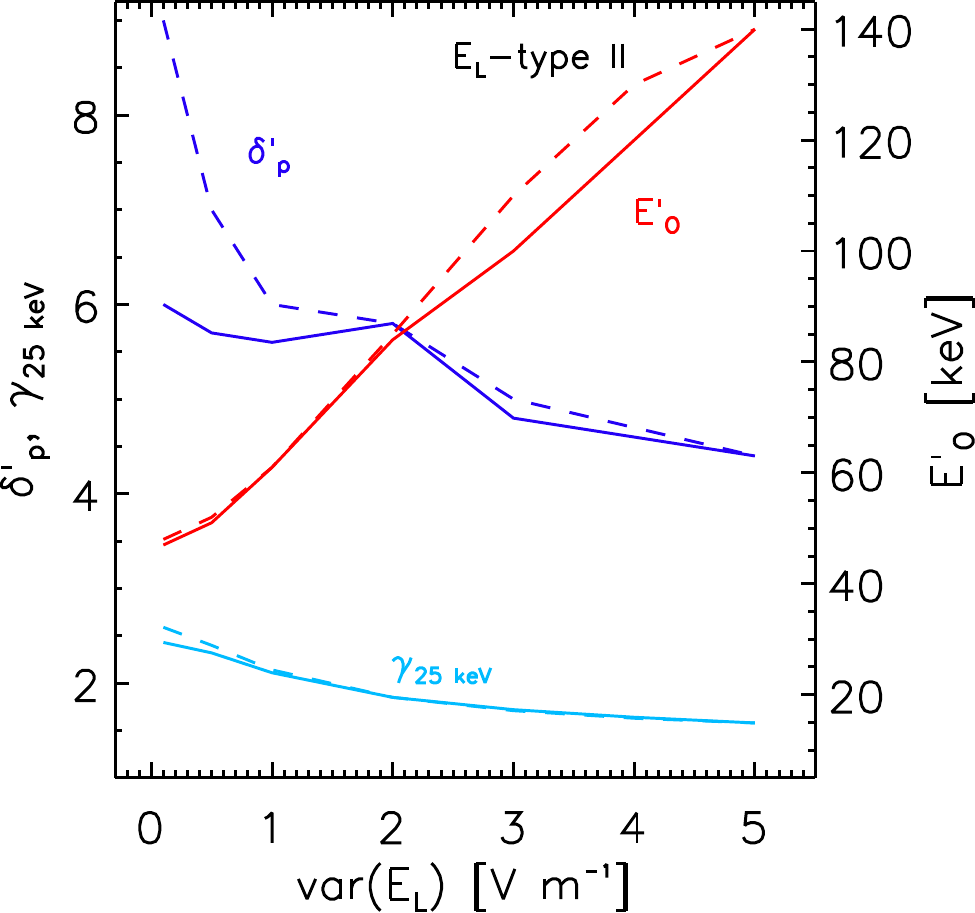}}
    \caption{LRTTM ${\bf E}_{\mathrm{L}}$-I ({\it left}) and  ${\bf
          E}_{\mathrm{L}}$-II ({\it right}) (for
        $\overline{E_{\mathrm{L}}}=0.1$~V\,m$^{-1}$)
        summary of calculated and fitted parameters of chromospheric
        bombardment for various values of
        $\mbox{var}({E_{\mathrm{L}}})$.
        {\it Top:} position of energy deposit maximum $s_{\max}$ and HXR intensity
        $I_{25\,\mathrm{keV}}$. {\it Bottom:} HXR spectral index $\gamma_{25\,\mathrm{keV}}$
        and fitted electron beam spectral index $\delta'_\mathrm{p}$ and
        low-energy cutoff $E'_0$. The solid and dashed lines
        denote $M^{\mathrm{FF}}$ and $M^{\mathrm{SU}}$, respectively.}
    \label{fig:LRTTM_b_new}
\end{figure}

\begin{figure*}[!t]
  \centerline{\includegraphics[width=8.cm]{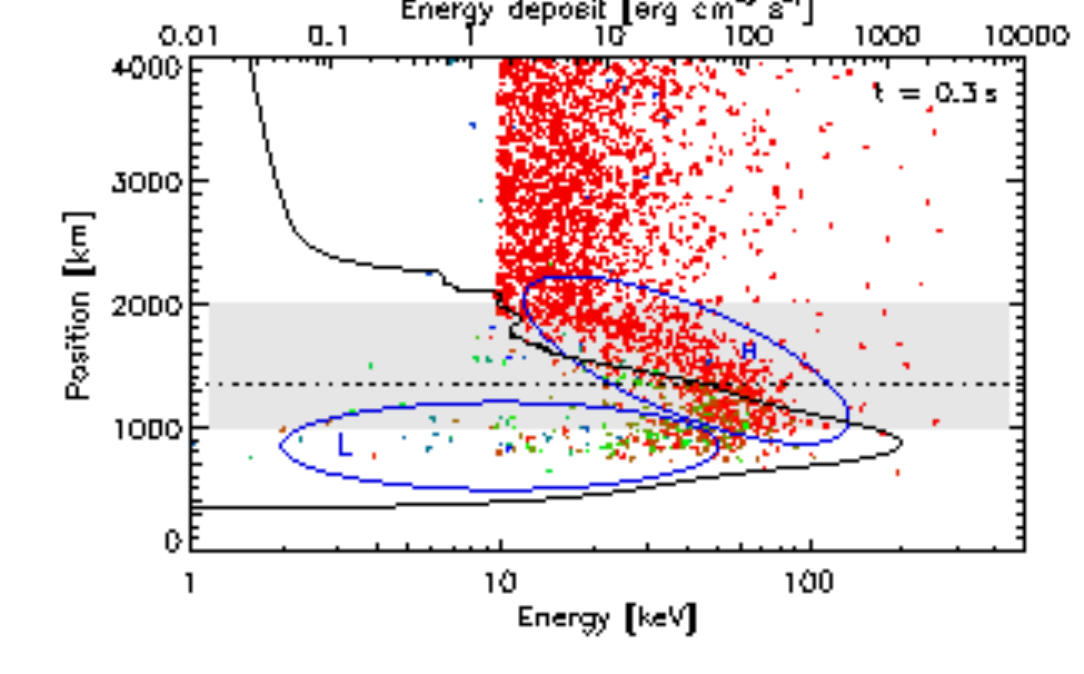}
              \includegraphics[width=8.cm]{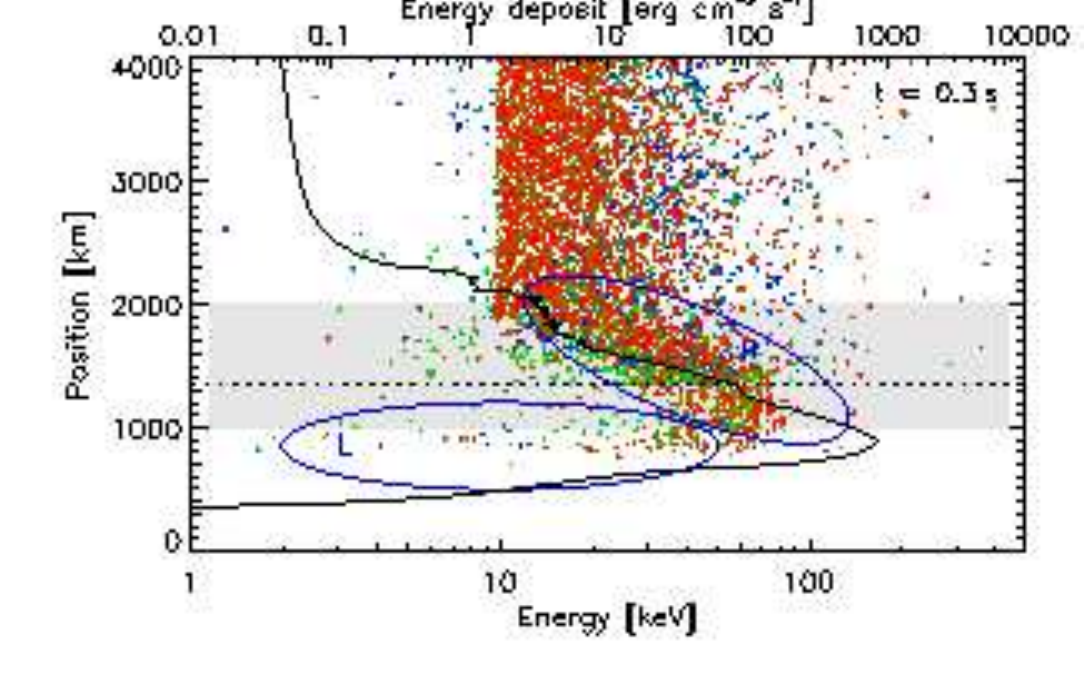}}
\vspace{-2em}
\centerline{\includegraphics[width=6cm]{fig_0.pdf}}
\caption{LRTTM ${{\bf E}_{\mathrm{L}}}$-II type distribution
functions of the non-thermal electron energies versus positions
with a colour coded $M(\mu)$ corresponding to
 $\overline{E_{\mathrm{L}}} = 0.1$~V\,m$^{-1}$ and
$\mbox{var}(E_{\mathrm{L}}) =0.5$~V\,m$^{-1}$ in the VAL C
atmosphere at time $t=0.3$~s after the  beam injection into the loop at its apex.
{\it Left:} $M^{\mathrm{FF}}$, {\it right:} $M^{\mathrm{SU}}$. The
solid lines indicate the instantaneous energy deposits corresponding to
$\mathcal{F}_{\mathrm{0}}/2=2.5\times10^{9}$~erg\,cm$^{-2}$\,s$^{-1}$,
the dotted horizontal lines the bottom boundary of the magnetic
mirror, the grey area the secondary re-acceleration region,
and the blue ellipses labelled L and H denote tails in the particle
distribution function. Only the vicinity of the footpoints is
displayed.}
\label{fig:13}
\end{figure*}

\begin{figure*}
  \centerline{\includegraphics[width=8.cm]{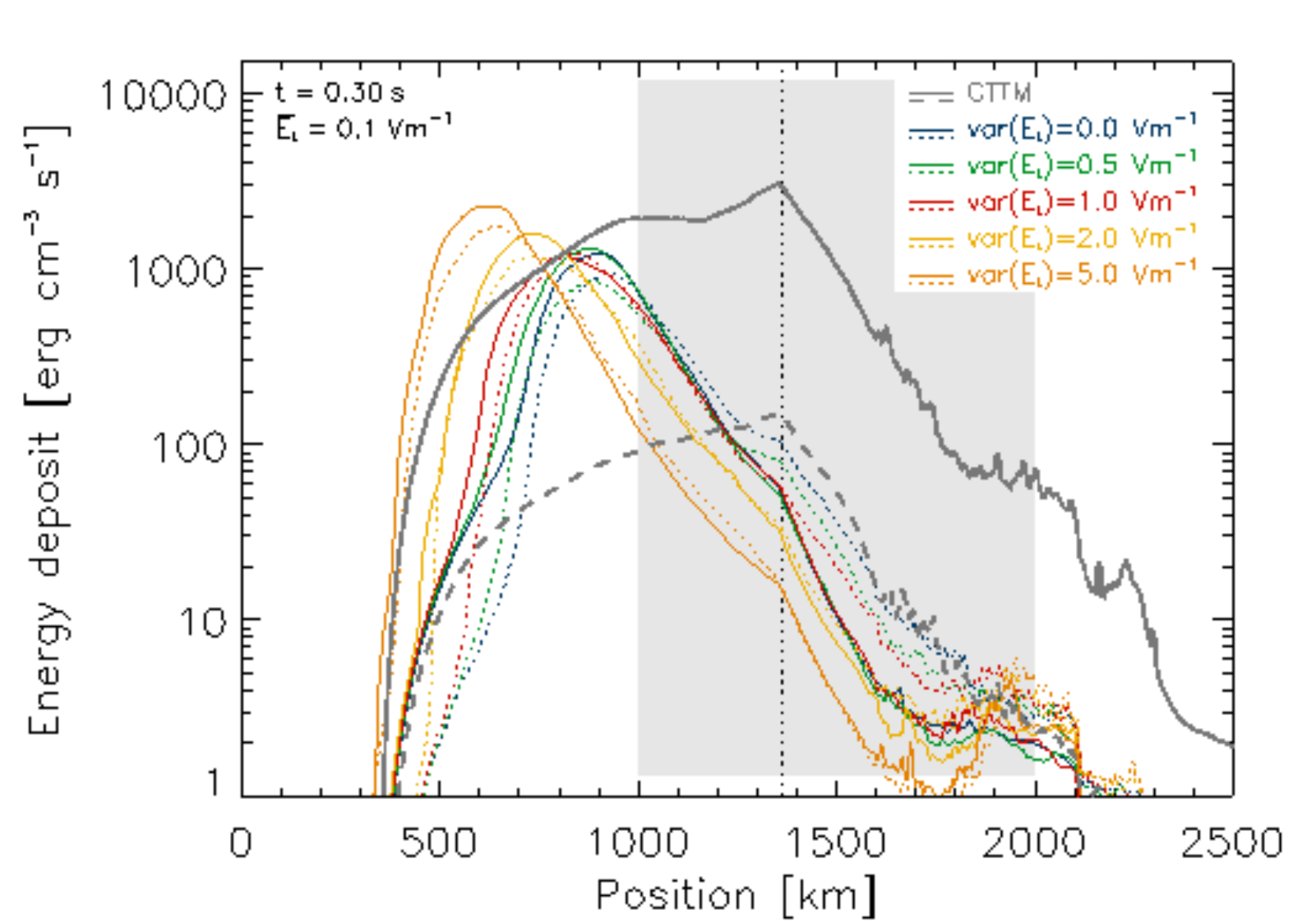}
              \includegraphics[width=8.cm]{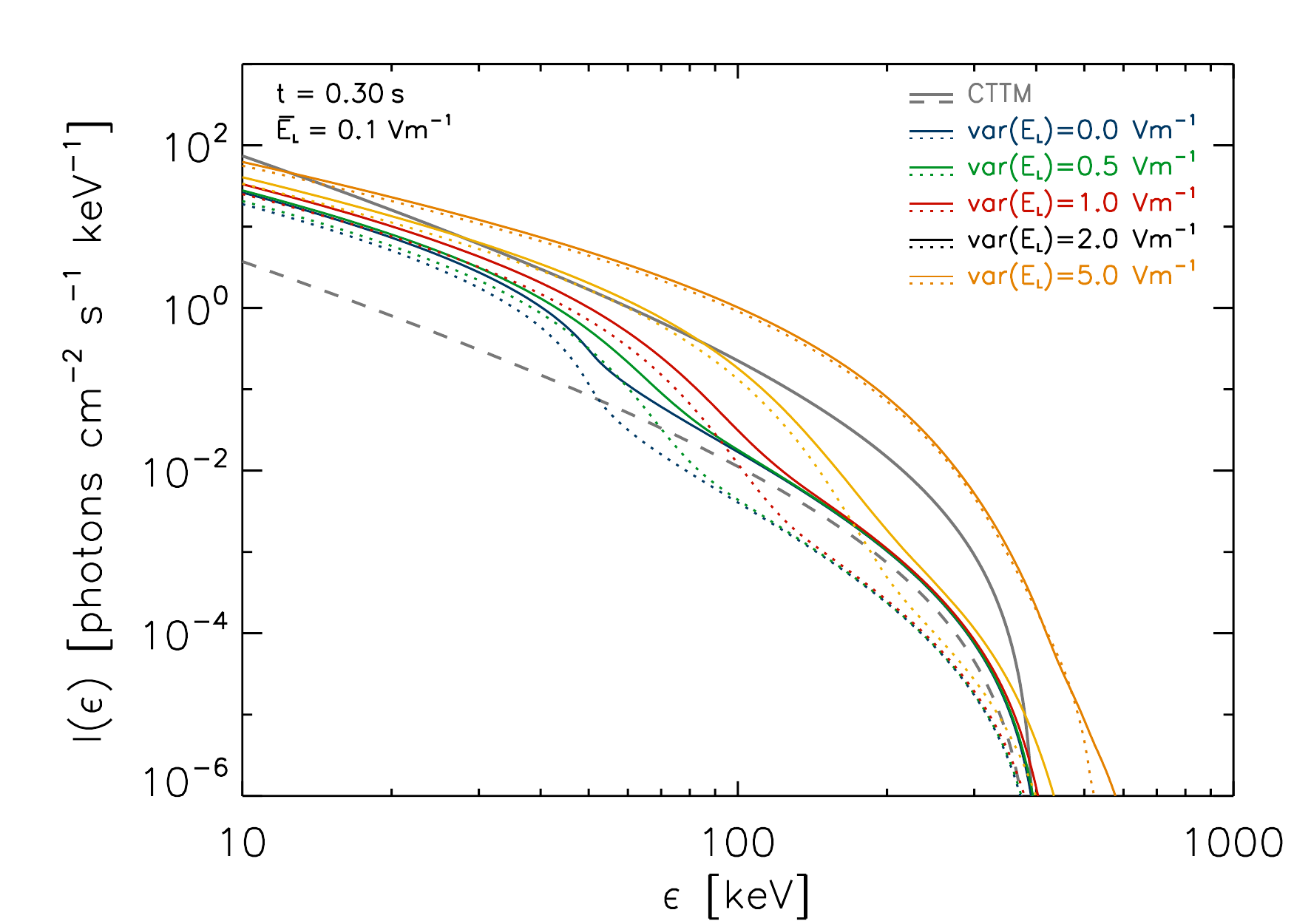}}
\caption{LRTTM ${{\bf E}_{\mathrm{L}}}$-II type
  instantaneous energy deposits ({\it left}) and HXR spectra
  ({\it right}) for the VAL~C  atmosphere at $t=0.3$~s. The solid
  ($M^{\mathrm{FF}}$) and dotted ($M^{\mathrm{SU}}$) blue,
  green, red, yellow, and orange lines correspond to
  $\overline{E_{\mathrm{L}}}=0.1$~V\,m$^{-1}$ and
  $\mbox{var}(E_{\mathrm{L}})=0.1, 0.5, 1, 2, 5~$~V\,m$^{-1}$, respectively
  and energy flux $\mathcal{F}_{\mathrm{0}}/2=2.5\times10^{9}$~erg\,cm$^{-2}$\,s$^{-1}$.
  The dashed and solid grey lines correspond to the CTTM
  with $\mathcal{F}_{\mathrm{0}}/2= 2.5\times10^{9}$ and
  $5\times10^{10}$~erg\,cm$^{-2}$\,s$^{-1}$, respectively.
  The dotted straight vertical line indicates the
  bottom boundary of the magnetic mirror, the grey area
  the secondary re-acceleration region. The HXR
  spectra are integrated over one half of the loop.}
\label{fig:14}
\end{figure*}

The stochastic field of $\overline{E_{\mathrm{L}}}=0.1$~V\,m$^{-1}$ and
$\mbox{var}(E_{\mathrm{L}})=0.5$~V\,m$^{-1}$ (see Fig.~\ref{fig:13})
practically ceases the formation of the low-energy tail of particles
located in the region between the upper boundary of the
re-acceleration region and the lower boundary of the magnetic mirror
found in the distribution functions corresponding to the CTTM, GRTTM,
and LRTTM ${{\bf E}_{\mathrm{L}}}$-I type (see Figs.~\ref{fig:4},
\ref{fig:6}, and \ref{fig:9}). It forms either for lower values of
$\overline{E_{\mathrm{L}}}$, which is too small to compensate for
the collisional energy losses of the electrons in the region above the
lower boundary of the magnetic mirror, or for greater values of
$\mbox{var}(E_{\mathrm{L}})$, when the interactions of beam electrons
with the stochastic component of ${{\bf E}_{\mathrm{L}}}$ lead to its formation.
On the other hand, a new tail of particles
is formed on energies from approximately 1 to 100~keV in the region
under the lower boundary of the re-acceleration region where the
re-accelerated particles are quickly thermalised (see the regions labelled
L).

The energy deposits and HXR spectra for $\overline{E_{\mathrm{L}}}=0.1$~V\,m$^{-1}$
and various values of $\mbox{var}(E_{\mathrm{L}})$ from 0 to 5~V\,m$^{-1}$
are plotted in Fig.~\ref{fig:14}, and the parameters
$\mathcal{E}_{\mathrm{ch}}$, $s_{\max}$, $I_{25\,\mathrm{keV}}$,
and $\gamma_{25\,\mathrm{keV}}$ are displayed in the left-hand and right-hand
panels of Figs.~\ref{fig:LRTTM_a_new} and \ref{fig:LRTTM_b_new},
respectively, and summarised in Table~\ref{tab:LRTTM}.
The general behaviour of  $\mathcal{E}_{\mathrm{ch}}$ and $s_{\max}$
is similar to the GRTTM of primary footpoint and LRTTM
${{\bf E}_{\mathrm{L}}}$-I type.
They are very sensitive to the static component $\overline{E_{\mathrm{L}}}$ of
the stochastic field and only moderately sensitive to the stochastic component
$\mbox{var}(E_{\mathrm{L}})$. Even for $\mbox{var}(E_{\mathrm{L}})=0$ and
$\overline{E_{\mathrm{L}}}=0.1$~V\,m$^{-1}$, there is an appreciable increase of
$\mathcal{E}_{\mathrm{ch}}$ ($3.6\times$ for the $M^{\mathrm{FF}}$ and
$5.5\times$ for the $M^{\mathrm{SU}}$ case) and a shift of $s_{\max}$ of
approximately 450~km towards the photosphere and substantial growth in HXR
production ($I_{25\,\mathrm{keV}}$ increases of by an order of magnitude for both
initial $\mu$-distributions relative to the CTTM  with an identical
initial flux). For the identical
value of $\overline{E_{\mathrm{L}}}$ and the maximum value of
$\mbox{var}(E_{\mathrm{L}}) =5$~V\,m$^{-1}$, the increase in
$\mathcal{E}_{\mathrm{ch}}$ is $5.5\times$ for the $M^{\mathrm{FF}}$ and
$10\times$ for the $M^{\mathrm{SU}}$ case, the shift of $s_{\max}$ towards the
photosphere of approximately 750~km (for both initial $\mu$-distributions), and
a substantial increase in $I_{25\,\mathrm{keV}}$ ($35\times$ for the
$M^{\mathrm{FF}}$ and almost $130\times$ for the $M^{\mathrm{SU}}$ case)
relative to the CTTM  with an identical initial flux. The power-law index
$\gamma_{25\,\mathrm{keV}}$ tends to harden with increasing
$\mbox{var}(E_{\mathrm{L}})$.

HXR spectra corresponding to the ${\bf E}_{\mathrm{L}}$-II type are
distinct from the previous ones. Here, two re-accelerating processes
are involved. The static component causes a significant increase of
spectra at deka-keV energies, up to $\sim$40~keV, and a steep double
break at energies above. Therefore, the corresponding fitted electron
flux spectrum assuming pure CTTM shows quite a steep $\delta_0'$ (see
Fig.~\ref{fig:LRTTM_b_new}, bottom right). Such a steep double break
is a consequence of a re-acceleration by a constant electric field. The
energy at which it appears is related to the length of the
re-acceleration region, i.e. the current sheet size. The larger the
size, the steeper the double break and the higher energies
at which it is located. The presence of the stochastic component
introduces another shift of the double break to higher energies,
likewise for the type I; as $\mbox{var}(E_{\mathrm{L}})$ increases, the double break
is less prominent. Consequently, $E_0'$ increases and $\delta_0'$
decreases (see Figs.~\ref{fig:LRTTM_b_new} and \ref{fig:14}).
When the stochastic component prevails,
i.e. $\mbox{var}(E_{\mathrm{L}})\ge 2$~V~m$^{-1}$,
the hard X-ray spectra are of similar spectral shape to the
${\bf E}_{\mathrm{L}}$-I model but more intense.

\section{Conclusions}
\label{sec:concl}

We studied modifications of the CTTM by considering two
types of secondary particle acceleration: GRTTM and LRTTM. In both
cases the re-acceleration takes place during the transport of
non-thermal particles, which are primarily accelerated in the corona.
According to  \citet{brown2009}, such a re-acceleration generally reduces
collisional energy loss and Coulomb scattering and increases the life-time
and penetration depth of particles.

In the case of GRTTM, the spatially varying direct electric field spreading along the
whole magnetic strand from first to second footpoint re-accelerates
the beam electrons towards the primary footpoint and decelerates
them towards the secondary footpoint, thus producing an asymmetric
heating of footpoints. The low electric plasma conductivity and
increased current density due to magnetic field convergence are
the key constraints for the functionality of this mechanism. The model
was studied for the mirror ratio $R_{\mathrm{m}}=5$ and current
densities $j\le6$~A~m$^{-2}$. Significant re-acceleration is present
for $j\gtrsim3$~A~m$^{-2}$, and for lower $j$ the model is similar
to CTTM. However, a question arises as to whether such current
densities are realistic. Although the current densities derived from magnetic
field observations are two orders of magnitude lower \citep{Guoetal13}, in the
magnetic rope, especially in their unstable phase at the beginning of the
flare, the current density in some filaments could reach these values: see
the processes studied in \citet{gordovskyy2011, gord2012, gord2013}.
On the other hand, a current filamentation also means a decrease in
the area where this re-acceleration can operate effectively.
Finally, the GRTTM model inherently introduces an asymmetry on
opposite sites of the magnetic rope. More
observations are needed to check that some asymmetrical X-ray sources
are caused by this effect.

Two types of  electric field were considered for LRTTM: a purely
stochastic field $\mbox{var}(E_{\mathrm{L}})\le 5$ ~V~m$^{-1}$
(${\bf E}_{\mathrm{L}}$-I type)  and a combination of
$\mbox{var}(E_{\mathrm{L}})$  and a static component
$\overline{E_{\mathrm{L}}}=0.1$~V~m$^{-1}$
(${\bf E}_{\mathrm{L}}$-II type). It has been shown that both
types of electric fields produce a substantial secondary
re-acceleration (${\bf E}_{\mathrm{L}}$-I type for
$\mbox{var}(E_{\mathrm{L}})\gtrsim0.5$~V~m$^{-1}$, ${\bf E}_{\mathrm{L}}$-II
type for all considered field parameters due to the static field component) with dominant
energy propagating towards the photosphere.

Generally in all presented models, HXR spectra gets flatter below
$\sim$30~keV and more intense on all energies as re-accelerating
fields increase. The flattening then corresponds to an increase in
the low-energy cutoff $E_0'$ of the fitted electron distribution. The
effect of flattening of HXR spectra below the low-energy cutoff can
be seen in \citet[][Fig.~1e]{brownkasp2008}. Extremely flat HXR
spectra (related to $E_0'\gtrsim 50$~keV) were obtained for GRTTM
of $j=6$~A~m$^{-2}$ and LRTTM
$\mbox{var}(E_{\mathrm{L}})\ge 2$~V~m$^{-1}$
(${\bf E}_{\mathrm{L}}$-I type). Such flat spectra or high values
of  $E_0'$ are not reported from the observation, therefore those $j$
and $\mbox{var}(E_{\mathrm{L}})$ could represent limiting values. In
addition, prominent double breaks at keV energies, present in the
${\bf E}_{\mathrm{L}}$-II cases, are not observed in HXR spectra. This
suggests that our model of a constant re-accelerating field over a
larger spatial scale, $\sim$1~Mm, is probably too simplistic.

For upper limit of model parameters, both models give similar results
in several aspects (although the values are probably extreme, at least
from the HXR signatures). At energies above 20~keV, the corresponding
HXR spectra are more intense than the spectrum of pure CTTM with $20\times$ higher
initial energy flux. GRTTM gives a comparable total chromospheric energy deposit. For the
LRTTM the total energy deposits reach only about 30\% of the latter value.
The re-acceleration also leads to spatial redistribution of the
chromospheric energy deposit with the bulk energy being deposited much
deeper into the chromosphere and into a narrower layer in comparison to the CTTM.
The heights of the energy-deposit maxima are
thus substantially shifted towards the photosphere (of $\approx$800~km
for both models). It is a consequence of the re-accelerating fields
pushing the non-thermal electrons under the magnetic mirror and
under the beam-stopping depth corresponding to the CTTM.
The height above the photosphere decreases with both the current
density for the GRTTM and with the mean value and variance of the stochastic
field for the LRTTM.  For the upper values of model parameters,
we obtained the heights of energy-deposit maxima as only approximately
600~km.
This is not far from the upper limits on heights of the flare
white-light sources  ($305\pm170$~km and $195\pm70$~km) found
from observations \citep{oliveros2012}.

To demonstrate how the secondary accelerating processes may lead to
artificially high CTTM input energy fluxes, we followed a standard
forward-fitting procedure for determining the injected electron
spectrum from an observed X-ray spectrum. Although the spectral
fitting does not take any re-acceleration into account, the fitted
$\mathcal{F}_0'$ agrees well (within $30\%$) with
$\mathcal{E}_\mathrm{ch}$ in all simulations. This value can differ
substantially from
the injected total energy flux, therefore the fitted total energy flux
(under assumption of pure CTTM) is related more to the energy
deposit of re-accelerated particles than to the injected
energy flux.

In general, both the considered models with secondary
re-acceleration, GRTTM and LRTTM, allow loosening the requirements
on the efficiency of coronal accelerator, thus decreasing
the total number of particles involved in the impulsive
phase of flares and the magnitude of the electron flux transported
from the corona towards
the photosphere, as needed to explain the observed HXR footpoint
intensities. These findings agree with the results obtained by
\citet{brown2009} and \citet{turkmani2006, turkmani2005}.

\begin{acknowledgements}
This work was supported by grants P209/10/1680 and P209/12/0103
of the Grant Agency of the Czech Republic. The research at the
Astronomical Institute, AS\v{C}R, leading to these results has received
funding from the European Commission's Seventh Framework Programme
(FP7) under the grant agreement SWIFT (project number 263340).
\end{acknowledgements}

\bibliographystyle{aa}
\bibliography{varady_et_al}

\end{document}